%% file: main1_1.tex
\documentclass[12pt]{iopart}
\usepackage{iopams}
\usepackage{citesort}
\usepackage{notes2bib}

\usepackage{graphicx}

\begin{document}

\title[Thermally induced subgap features in cotunneling spectroscopy of a carbon nanotube]{Thermally induced subgap features in the cotunneling spectroscopy of a carbon nanotube}

\input authors.tex

\begin{abstract}
We report on nonlinear cotunneling spectroscopy of a carbon nanotube quantum dot coupled to Nb superconducting contacts. Our measurements show rich \textit{subgap} features in the stability diagram which become more pronounced as the temperature is increased. Applying a transport theory based on the Liouville-von Neumann equation for the density matrix, we show that the transport properties can be attributed to processes involving sequential as well as elastic and inelastic cotunneling of quasiparticles thermally excited across the gap. In particular, we predict thermal replicas of the elastic and inelastic cotunneling peaks, in agreement with our experimental results.
\end{abstract}

%
%
%
%
%

\section{Introduction}
Due to proximity effects, a hybrid device made of a superconductor coupled to a mesoscopic normal conductor allows to study a wide range of quantum phenomena. In particular, in the Coulomb blockade regime these include supercurrent transport carried by Cooper pairs~\cite{glaz89,bas99,roz01,Doh05,dam06,Jar06}, coherent electron transport in terms of multiple Andreev reflections~\cite{Sch01,Bui03,andersen11,deon11}, as well as quasiparticle transport~\cite{levy97,GoLo04,dam06,eich07,grove09,dirk09,fran10,Pfaller13,Gaass14}. Andreev reflections lead to subgap structures with steps at bias voltage $2\Delta/ne$ ($n\in \mathbb{N}^+$) in the current-voltage characteristics~\cite{levy97,joh99,Bui03,eich07,andersen11,deon11,gunel12}, which are smeared out by increasing the temperature~\cite{eich07,deon11}. In contrast, temperature favors quasiparticle transport, as it increases the probability of thermal activation of quasiparticles across the gap. The emergence of a zero bias peak inside the Coulomb diamond by 
increasing temperature~\cite{eich07,deon11} was explained in terms of resonant tunneling~\cite{levy97} of thermal quasiparticles. Recently, the additional possibility to observe transport features due to sequential tunneling of thermally excited quasiparticles has been theoretically proposed in Ref. \cite{Pfaller13} and experimentally confirmed in Ref. \cite{Gaass14}. Such processes lead to thermal resonance lines within the Coulomb blockade region, parallel to the Coulomb diamond edges. Cotunneling processes due to quasiparticles, however, have so far only been reported for bias voltages above the superconducting energy gap~\cite{grove09,dirk09}. In this work we present measurements in complete agreement with theoretical predictions on thermally excited quasiparticle transport in the cotunneling regime.

Cotunneling is a transport process in which the QD is either excited (inelastic cotunneling), or kept in the same state as the initial state (elastic cotunneling), by means of tunneling events to an intermediate virtual state. Thus, for the inelastic case a bias threshold corresponding to the excitation energy is required to enable charge transfer~\cite{averin90}. In contrast to sequential tunneling processes, cotunneling in lowest order is expected to be independent of the gate voltage.

We report on elastic and inelastic cotunneling spectroscopy on individual carbon nanotube (CNT) devices coupled to Nb superconducting leads. In the low temperature limit transport theory predicts for a CNT quantum dot superconductivity enhanced transport features at bias voltages $\pm2\Delta/e$ and $\pm(2\Delta+\delta_m)/e$ due to elastic and inelastic cotunneling of quasiparticles, respectively~\cite{grove09}. Here \{$\delta_m$\} is the set of excitation energies of the CNT from an N particle ground state. With increasing temperature, we predict and observe the appearance of elastic and inelastic cotunneling features in the subgap region (i.e. for bias voltage amplitudes smaller than $2\Delta/e$) due to thermally excited quasiparticles. In particular, the emergence of a zero-bias peak, corresponding to the thermal replica of the elastic cotunneling resonance, is expected. Our theoretical predictions are in good quantitative agreement with our experimental findings.

\begin{figure}[tb]
\centering
\includegraphics[width=\columnwidth]{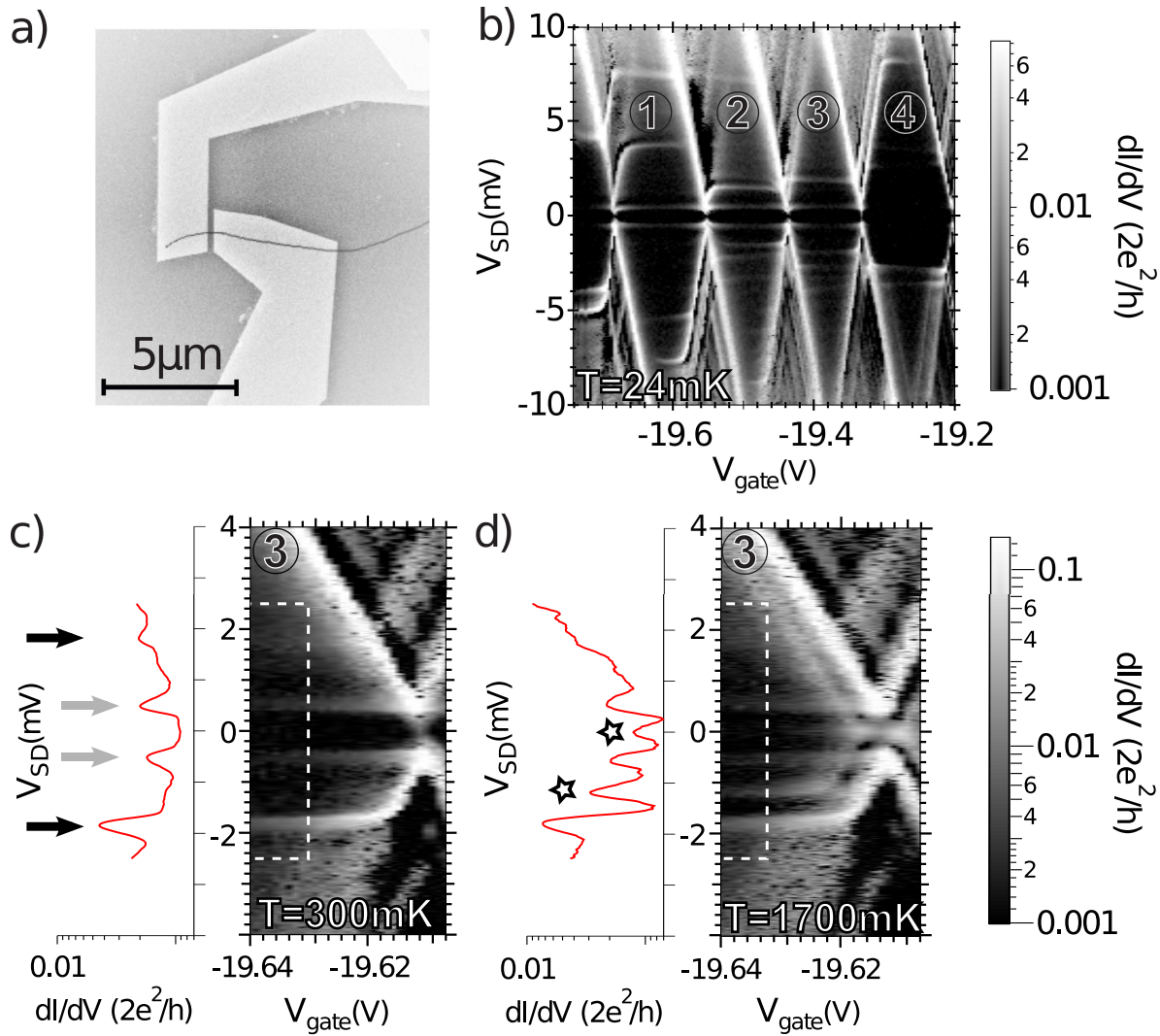}
\caption{(a) Scanning electron micrograph of the device \textsf{A}. The gray line indicates the approximated location of the nanotube (not visible itself). (b) Differential conductance at $T=24\,$mK as function of bias voltage and back gate voltage. (c)(d) Zoom into the Coulomb blockade region of the third Coulomb diamond for temperatures $T=300\,$mK and $T=1700\,$mK, respectively. The dashed white box corresponds to the range of gate voltages over which is averaged to obtain the differential conductance curve shown on the left side of each figure.
}\label{fig:sampleA}
\end{figure}

Individual single wall carbon nanotubes were grown on a highly p-doped Si/SiO$_2$ substrate by chemical vapor deposition~\cite{kong98}. The substrate acting as a global back gate is used to tune the electron occupation of the CNT. The source and drain electrodes were patterned on an individual single wall carbon nanotube by standard electron beam lithography and lift-off techniques. Here we report on measurements on two distinct samples. For sample \textsf{A}, Fig.~\ref{fig:sampleA}, electrodes made of $3\,$nm Pd and $45\,$nm sputtered Nb with a spacing between electrodes of the order of $300\,$nm were used (see Fig.~\ref{fig:sampleA}(a)); for sample \textsf{B}, Fig.~\ref{fig:amit} in the appendix, a metalization of $3\,$nm Pd and $60\,$nm sputtered Nb with a contact spacing of the order of $430\,$nm was applied (see Fig.~\ref{fig:amit}(a)). In order to perform four-point measurements and as a resistive on-chip element, each superconducting electrode was connected to two leads made of AuPd to damp 
oscillations at the plasma 
frequency of the Josephson junction~\cite{pall08,mart89}. Low temperature electrical transport measurements were performed inside a 3He/4He dilution refrigerator with a base temperature of $25\,$mK.

In both samples we observe regular CB diamonds over a large gate voltage range. Signatures of four-fold periodicity are observed in the measured gate range only for Sample \textsf{A}. Figs.~\ref{fig:sampleA}(b) and \ref{fig:amit}(b) show the high resolution measurements for selected gate range for contacts in the superconducting state at temperature $T=25\,$mK and $30\,$mK, respectively. In both samples lines of high conductivity are observed well inside the Coulomb diamonds; all these lines are horizontal, independent of gate voltage. To clearly identify them we restrict the gray scale for the differential conductance below the maximum conductance. Fig.~\ref{fig:sampleA}(c) shows a zoom corresponding to the region inside the diamond denoted \textcircled{\bf\footnotesize{3}} in Fig.~\ref{fig:sampleA}(b)~\bibnote{The discrepancy in gate voltage range between Fig.~\ref{fig:sampleA}(b) and Figs.~\ref{fig:sampleA}(c),(d) is caused by a long-time scale drift of all Coulomb blockade features. Sequential tunneling 
features of this data 
set have 
already been discussed in Ref. \cite{Gaass14}}. Horizontal lines are clearly visible and indicated by arrows in the conductance curve.

One set of line occurs at bias voltage $V_\textsuperscript{SD}\sim\pm 0.52\,$mV (gray arrows). We ascribe it to elastic cotunneling processes at $V_\textsuperscript{SD}=\pm2\Delta/e$. We extract $\Delta\sim0.26\,$meV for our superconducting film, compared to the expected value of $\Delta=1.5\,$meV for bulk Nb. The mismatch of about a factor of five has already been reported in similar Nb-based devices~\cite{kulm72,may72,grove09,Kum14}. The reason for the gap reduction is still an open question. Possible explanations are the formation of niobium oxide, the thin composite of Nb and Pd, or the contamination of the lower Nb interface. For the deposited Nb/Pd strip a critical temperature of about 8 K was measured, where the resonant features remain present up to temperatures of about 4 - 5 K. Thus the transition temperature of the thin film is comparable to bulk Nb, that is in contrast to the observed small value of $\Delta$ and the BCS-relation $\Delta=1.76k_BT$. The inelastic part of the cotunneling spectra reveals excitations of the CNT quantum dot. Our data show a broad inelastic feature at a distance $\delta=1.3\,$meV from the elastic line (black arrows). From additional stability diagrams for sample \textsf{A}, recorded at higher temperature and finite magnetic field to suppress superconductivity, we extract a charging energy $E_\textsuperscript{C} \simeq  15\,$meV, implying $E_\textsuperscript{C}/\Delta\sim50$ for sample \textsf{A}. Similarly, from the elastic and inelastic line, we can also extract $\Delta\sim0.23\,$meV and $\delta\sim0.11\,$meV for sample \textsf{B}. From additional stability diagrams in a regime in which superconductivity is largely suppressed, we identify a smaller charging energy $E_\textsuperscript{C}\simeq3.2\,$meV. The two samples have roughly the same superconducting gap $\Delta$ but differ in the charging energy $E_\textsuperscript{C}$, leading to different transport regimes. In both samples charging effects and the small coupling strength $\hbar\Gamma<\Delta$ suppress Andreev processes, such that current is carried by quasiparticles. In sample \textsf{A} the large charging energy further suppresses multiple quasiparticle processes. Thus the transport is dominated by sequential and cotunneling events. For sample \textsf{B} a simple description in terms of resonant tunneling of quasiparticles~\cite{levy97} may be conceived.

As the temperature is increased new horizontal lines are observed. In sample \textsf{A} the novel lines arise for temperatures above $T\approx 600\,$mK at zero-bias and at bias voltage $V_\textsuperscript{SD}=\pm\delta/e$. Fig.~\ref{fig:sampleA}(d) shows the same gate region as in Fig.~\ref{fig:sampleA}(c) but now for the temperature $T=1.7\,$K. The additional lines, marked by stars, become more and more pronounced with increasing temperature. Andreev reflections do not give an explanation for the thermal behavior of such transition lines~\cite{Sch01,Bui03,Doh05,Jar06,Kum14}. Also the Kondo effect cannot be the reason for the resonant peak at zero bias, as it has an opposite thermal behavior~\cite{bui02,sia04,Cle06,Kim13,Lee14,Chang13}.

The feature of a zero-bias conductance peak is also supported by sample $\textsf{B}$, as shown in Fig.~\ref{fig:amit}(c) in the appendix. The bias trace is taken in the middle of the Coulomb blockade valley at gate voltage $V_\textsuperscript{gate}\approx-11.71\,$V. Upon increase of the temperature, one observes a rising conductance peak at zero bias, and pairs of symmetrically displaced elastic and inelastic cotunneling peaks at finite bias. The feature at bias voltage $V_\textsuperscript{SD}=\pm\Delta/e$ and the thermal zero bias peak resemble data already reported in Refs. \cite{eich07,deon11}. In analogous fashion, we expect them to be reproducible within the simple resonant model of Ref. \cite{levy97}. The more complex behavior of sample \textsf{A}, where several cotunneling and sequential lines are observed within the CB diamond, clearly goes beyond the capability of the simple resonant picture that exclude Coulomb interaction. As shown below, a full transport theory including all tunneling processes 
up to 
second order in the coupling strength $\hbar\Gamma$ to the leads can capture the experimental behavior to high detail.

\section{Transport theory for S-CNT-S junctions}
To understand the experimental observations, we consider a minimal model for a CNT quantum dot connected to two BCS-type superconducting leads. For the back-gated CNT we consider a single longitudinal mode incorporating orbital, $m$, and spin, $\sigma$, degrees of freedom. Coulomb interaction effects are considered within a constant interaction model, with $U$ being the charging energy. The quadruplet CNT Hamiltonian thus reads
\begin{equation}
\hat H_\textsuperscript{CNT}=\sum_{m\sigma} E_{m\sigma}\hat d^\dagger_{m\sigma}\hat d_{m\sigma}+\frac{U}{2}\hat N(\hat N-1)-\alpha eV_\textsuperscript{gate}\hat N,
\end{equation}
where $\hat N$ is the charge number operator of the dot and $\alpha$ a conversion factor for the gate voltage. Finally, $E_{m\sigma}=\epsilon_d+\frac{1}{2}m\sigma\delta$ (with $m=\pm1$, $\sigma=\pm1$), where $\delta$ accounts for the breaking of the fourfold degeneracy of a longitudinal mode with energy $\epsilon_d$ due to spin-orbit interaction and valley mixing~\cite{laird14}.

The BCS superconducting leads are described by a conventional pair-interaction Hamiltonian on a mean-field level with respect to an offset energy $E_l^0$: 
\begin{equation}
 \hat H_l=E_l^0+\sum_{\vec k\sigma}E_{l\vec k}\hat\gamma^\dagger_{l\vec k\sigma}\hat\gamma_{l\vec k\sigma}+\mu_l\hat N_l.
\end{equation}
It can be obtained by means of a particle conserving Bogoliubov-Valatin transformation~\cite{Bog58,Val58}
\begin{eqnarray} 
 \hat c^\dagger_{l\vec k\sigma}&=&u_{l\vec k}\hat\gamma^\dagger_{l\vec k\sigma}+\sigma v^*_{l\vec k}\hat S^\dagger_l\hat\gamma_{l-\vec k\bar\sigma},\nonumber \\*
 \hat c_{l\vec k\sigma}&=&u^*_{l\vec k}\hat\gamma_{l\vec k\sigma}+\sigma v_{l\vec k}\hat S_l\hat\gamma^\dagger_{l-\vec k\bar\sigma},
\end{eqnarray}
for the leads' electron creation and annihilation operators $\hat c^\dagger_{l\vec k\sigma}$ and $\hat c_{l\vec k\sigma}$, respectively. The electron operators are represented in terms of quasiparticle operators $\hat\gamma^{(\dagger)}_{l\vec k\sigma}$ and of Cooper pair operators $\hat S^{(\dagger)}_l$ with the corresponding prefactors $u^{(*)}_{l\vec k}$ and $v^{(*)}_{l\vec k}$~\cite{Jos62,Bar62}. Furthermore, the quasiparticles have an excitation energy  $E_{l\vec k}=\sqrt{(\epsilon_{\vec k}-\mu_l)^2+\Delta^2}$ measured with respect to the electrochemical potential $\mu_l$. Finally, the BCS gap is defined by
\begin{eqnarray}
 \Delta\equiv |V|\sum_{\vec k}\left\langle \hat S^\dagger_l\hat c_{l-\vec k\downarrow}\hat c_{l\vec k \uparrow}\right\rangle\label{gap},
\end{eqnarray}
where $|V|$ characterizes the interaction potential between a pair of electrons.

The connection with the superconducting leads is realized by a single-particle tunneling Hamiltonian $\hat H_{T,l}=T_l\sum_{\vec k \sigma m}\left(\hat d^\dagger_{m\sigma}\hat c_{l\vec k\sigma}+h.c.\right)$ where, for the sake of simplicity, the tunnel coefficient $T_l$ of lead $l$ is considered to be spin, wave vector and valley independent. The tunnel coupling strength can then be defined as $\hbar\Gamma_l\equiv2\pi|T_l|^2\sum_{\vec k}\delta(\omega-\epsilon_{\vec k})$, which is assumed to be energy independent.

We describe the time evolution of the system with the generalized master equation~\cite{Blu12}:
\begin{eqnarray}
   \dot{\hat\rho}_\textsuperscript{red}(t)=-\frac{i}{\hbar}\left[\hat H_\textsuperscript{CNT},\hat\rho_\textsuperscript{red}(t)\right]+\int^t_{t_0}d\tau\hat K(t,\tau)\hat\rho_\textsuperscript{red}(\tau),
\end{eqnarray}
for the dynamics of the reduced density operator $\hat\rho_\textsuperscript{red}$. This (still exact) equation allows a systematic perturbation expansion of the kernel superoperator $\hat K(t,\tau)$ in powers of the coupling strength $\hbar\Gamma$~\cite{weymann05,Koller10}. In the steady state limit and charge conserved regime the master equation can be simplified further by applying the Laplace transform $f(\lambda)\equiv\int^\infty_0 d\tau'\,e^{-\lambda \tau'}f(\tau')$ and its properties:
\begin{eqnarray}
 0=-\frac{i}{\hbar}\sum_{\chi_i\chi'_i}\delta_{\chi_i\chi_f}\delta_{\chi'_i\chi'_f}(E_{\chi_i}-E_{\chi'_i})\rho_{\chi_i\chi'_i}+\sum_{\chi_i\chi'_i}K^{\chi_i\chi'_i}_{\chi_f\chi'_f}\rho_{\chi_i\chi'_i},
\end{eqnarray}
with $K^{\chi_i\chi'_i}_{\chi_f\chi'_f}\equiv \langle \chi_f|\hat K(\lambda=0^+)[|\chi_i\rangle\langle\chi'_i|]|\chi'_f\rangle$ and $\rho_{\chi_i\chi'_i}\equiv\langle\chi_i|\hat\rho_\textsuperscript{red}(t\to\infty)|\chi'_i\rangle$. The matrix elements are evaluated in the basis $\{|\chi\rangle\}$ of the eigenstates of the Hamiltonian $\hat H_\textsuperscript{CNT}$. Noticeably, each term in the perturbation expansion of $K^{\chi_i\chi'_i}_{\chi_f\chi'_f}$ can be represented in a diagrammatic language in which simple rules exist to directly obtain the corresponding analytical expression. In Ref. \cite{gov08} these rules are derived and discussed in detail for the case of hybrid S-QD-S nanostructures.

An expression for the steady state current in terms of a perturbative expansion can be obtained in the same way. In particular, the net current of lead $l$ is described by
\begin{eqnarray}
 I_l(t\to\infty)=e\sum_{\chi_f}\sum_{\chi_i\chi'_i}(K_{I_l})^{\chi_i\chi'_i}_{\chi_f\chi'_f}\rho_{\chi_i\chi'_i}.
\end{eqnarray}
In the charge conserved regime the reduced density matrix $\rho_{\chi_i\chi'_i}$ is block diagonal (see appendix D). Thus the kernel element $K^{\chi_i\chi'_i}_{\chi_f\chi'_f}$ up to second order also represents the physical rate for processes transferring 0, 1, or 2 charge(s), depending on the charge difference between the states $|\chi_i\rangle$ and $|\chi_f\rangle$.

The problem of non-equilibrium hybrid superconducting-quantum dot junctions with an applied bias voltage is intrinsically time dependent. This can lead to time-dependent harmonic contributions to the stationary current associated to Andreev tunneling~\cite{andersen11}. However, in the charge conserved regime considered in this work, these harmonics are absent, and hence $\dot{\hat\rho}_\textsuperscript{red}(t)\to 0$ at long times. This is because the expectation values $\langle\hat c^\dagger_{l\vec k\sigma}(t)\hat c^\dagger_{l'\vec k'\sigma'}(\tau)\rangle$ and $\langle\hat c_{l\vec k\sigma}(t)\hat c_{l'\vec k'\sigma'}(\tau)\rangle$ vanish since they break the conservation of total charge. Let us emphasize that, according to Eq.~(\ref{gap}), we still have a finite superconducting gap and superconducting features (see appendix C for a detailed discussion).

Thermally assisted quasiparticle transport has yet only been discussed in the context of sequential~\cite{Pfaller13,Gaass14} and resonant~\cite{deon11,eich07} tunneling. Responsible for the energy distribution of the fermionic quasiparticles is beside the BCS density of states (DOS) also the Fermi function. For high enough temperatures the Fermi function is thermally smeared, in the sense that quasiparticles can also occupy the high energy branch of the DOS and thus can contribute to an additional transport channel.
\begin{figure}[tb]
\centering
\includegraphics[width=.9\columnwidth]{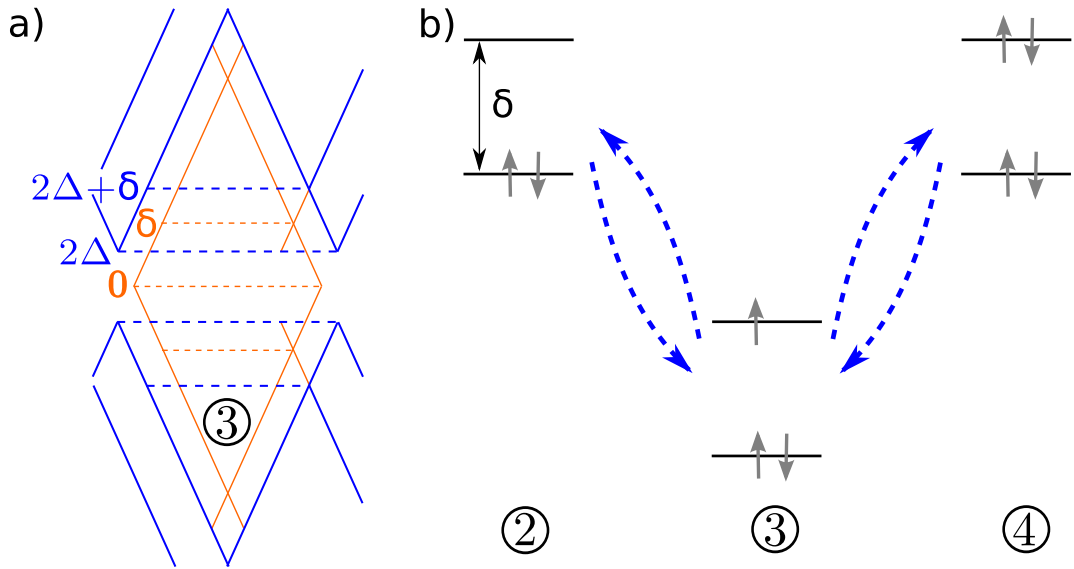}
\caption{a) Theoretically expected transition lines in the stability diagram of a CNT for one specific Coulomb diamond. Solid and dashed blue lines correspond to standard sequential tunneling and cotunneling processes, respectively. The thermal replicas of these transition lines are shown as solid and dashed orange lines. b) Many-body spectrum of the 2, 3 and 4 electron subspace for a gate voltage corresponding to the center of the Coulomb diamond \textcircled{\bf\scriptsize{3}}. The tunneling events contributing to the elastic cotunneling lines are shown.}\label{fig:thermal}
\end{figure}
In the sequential tunneling regime, this gives rise to thermal replicas of the sequential tunneling transitions displaced by $\pm4\Delta/e$ in bias voltage (solid orange lines in Fig.~\ref{fig:thermal}(a)). When cotunneling processes are also taken into account, the number of expected thermal lines is largely increased, as sketched in Fig.~\ref{fig:thermal}(a). In the figure we restrict us to the exemplary Coulomb diamond denoted \textcircled{\bf\footnotesize{3}}. Gate-dependent lines, induced by sequential processes, can be clearly distinguished from gate independent cotunneling induced lines. Blue solid and dashed lines are transitions which are due to ``standard'' sequential tunneling and cotunneling processes, respectively, i.e., contributions that are also present at low temperatures. Orange solid and dashed lines, in contrast, are due to thermally excited quasiparticles. Hence, they are present only at large enough temperatures.

As already mentioned, standard elastic cotunneling lines are expected at bias $V_\textsuperscript{SD}=\pm2\Delta/e$, and the inelastic cotunneling features occur at a bias $V_\textsuperscript{SD}=\pm(2\Delta+\delta)/e$, reflecting the excitation energy $\delta$. Fig.~\ref{fig:thermal}(b) visualizes the elastic cotunneling events in the many-body spectrum where the 3-particles ground-state is used as reference energy. Choosing the center of diamond \textcircled{\bf\footnotesize{3}}, corresponding to a certain gate voltage, the 2-particles and the 4-particles ground-state have the same energy. Thus, transitions from the 3-particles ground-state to the 2-particles ground-state and backwards have the same probability as those from the 3-particles ground-state to the 4-particles ground state and backwards, leading to elastic cotunneling. As shown below, thermal excitation of the lead quasiparticles yields thermal replicas at a bias $2\Delta/e$ smaller than for standard cotunneling features. We thus predict, in 
particular, the emergence 
of a cotunneling line at zero bias, being the thermal replica of the standard elastic lines at $\pm2\Delta/e$.

\begin{figure}[tb]
\centering
\includegraphics[width=\columnwidth]{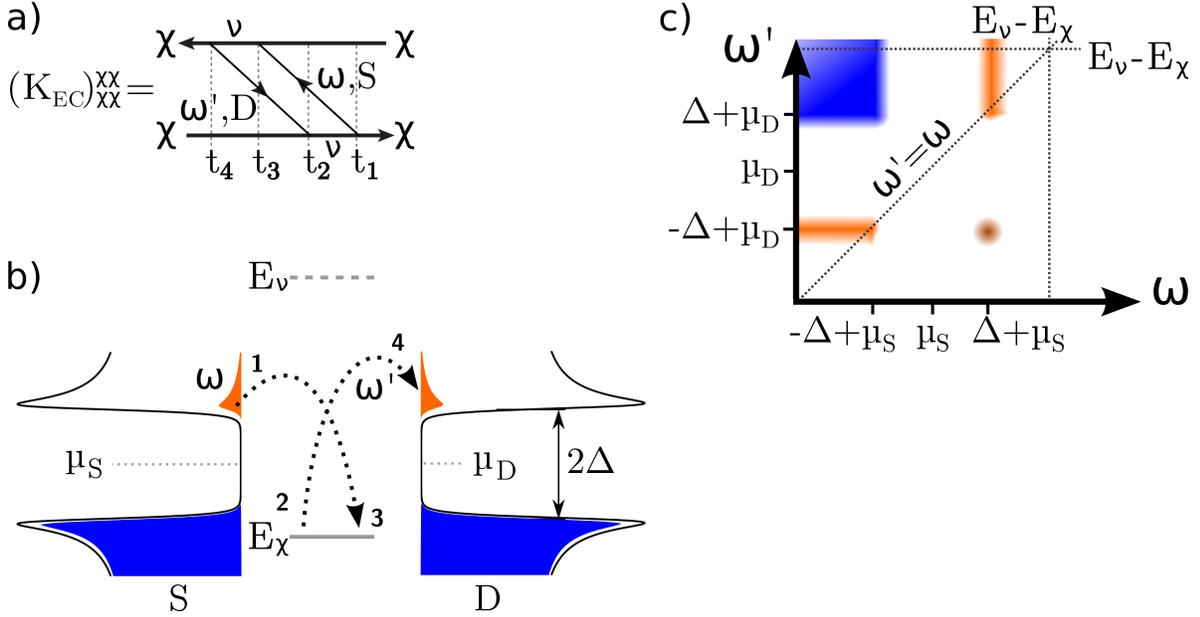}
\caption{(a) Exemplary diagrammatic representation of one main contribution to elastic cotunneling. (b) Energy-DOS diagram explaining the transport mechanism for thermally assisted elastic cotunneling. The time ordering of the tunnel processes has the same declaration as in the diagram (a). A measurable elastic cotunneling current is observed if thermally occupied quasiparticle states in the source are simultaneously aligned with empty quasiparticle states in the drain. (c) Integrand of Eq.~(\ref{eqcot}) for the parameter regime of figure (b). Blue corresponds to the low temperature parameter regime $T\ll\Delta/k_B$ where the product of Fermi functions and density of states is finite. Orange represents the area where the product has to be taken into account for higher temperatures $T<\Delta/k_B$. }\label{fig:cotunnel}
\end{figure}

One exemplary contribution to elastic cotunneling in the diagrammatic language is shown in Fig.~\ref{fig:cotunnel}(a).
Using the diagrammatic rules~\cite{gov08,Koller10} the analytic expression is given by the kernel element
\begin{eqnarray}
 (\hat K_\textsuperscript{EC})^{\chi\chi}_{\chi\chi}&\equiv&-i\hbar\Gamma_\textsuperscript{S}\Gamma_\textsuperscript{D}\sum_\nu\int \frac{d\omega}{2\pi} \frac{d\omega'}{2\pi} D_\textsuperscript{S}(\omega,\Delta)D_\textsuperscript{D}(\omega',\Delta) \nonumber \\* 
&&\times\frac{f_\textsuperscript{S}(\omega)(1-f_\textsuperscript{D}(\omega'))}{(-\omega+\delta E+i0^+)(\omega'-\omega+i0^+)(\omega'-\delta E+i0^+)} \nonumber \\*
&\equiv&-\frac{i}{\hbar}\Gamma_\textsuperscript{S}\Gamma_\textsuperscript{D}\sum_\nu\int \frac{d\omega}{2\pi} \frac{d\omega'}{2\pi}\,I(\omega,\omega'),\hspace{4.8em}\label{eqcot}
\end{eqnarray}
including  $f_l(\omega)\equiv 1/[\exp((\omega-\mu_l)/k_BT)+1]$, the DOS $D_l(\omega,\Delta)\equiv\sqrt{\frac{(\omega-\mu_l)^2}{(\omega-\mu_l)^2-\Delta^2}}$ $\times\Theta(|\omega-\mu_l|-\Delta)$, and the energy difference $\delta E=E_\nu-E_\chi$ between the energy $E_\nu$ of the virtual dot state $|\nu\rangle$ and $E_\chi$ of the dot state $|\chi\rangle$. Notice that in the example of Fig.~\ref{fig:cotunnel}(a) the state $|\nu\rangle$ has one unit of charge more than state $|\chi\rangle$. The charges entering and leaving the dot carry the energies $\omega$ and $\omega'$, respectively. An analysis of the double integral shows that, at low temperatures, it gives one pronounced contribution only in the case 
$V_\textsuperscript{SD}\ge2\Delta/e$ (see appendix E). The bias threshold $V_\textsuperscript{SD}=2\Delta/e$ corresponds to the resonant case in which the highest occupied quasiparticle states in the source are aligned with the lowest empty quasiparticle states in the drain, such that elastic cotunneling onto and out of the CNT is possible. However, at higher temperatures thermally excited quasiparticles enable cotunneling transport also at zero bias. This mechanism is visualized in Fig.~\ref{fig:cotunnel}(b), where the numbers \textsf{1}, \textsf{2}, \textsf{3}, \textsf{4} correspond to the tunneling events occurring at times $\tau\equiv t_1\le t_2 \le t_3 \le t_4\equiv t$ shown in Fig.~\ref{fig:cotunnel}(a). As seen in Fig.~\ref{fig:cotunnel}(b), if the thermally occupied quasiparticle states of the source are in resonance with the unoccupied quasiparticle states of the drain, elastic cotunneling through the dot can occur also at zero bias. The tunneling rate $\Gamma_\textsuperscript{EC}^{\chi\to\chi}\equiv2\textnormal{Re}(\hat K_\textsuperscript{EC})^{\chi\chi}_{\chi\chi}$ for such a process is given by the expression in Eq.~(\ref{eqcot}) adding the hermitian conjugated.

Mathematically, the condition for the onset of elastic cotunneling can be obtained from the analysis of the integrand $I(\omega,\omega')$ of Eq.~(\ref{eqcot}). This integrand is schematically depicted in Fig.~\ref{fig:cotunnel}(c) for the case of zero bias and $\Delta+\mu_\textsuperscript{S/D}\ll\delta E$, such that the system is in the Coulomb blockade regime and no sequential transport occurs. Due to the product $D_\textsuperscript{S} D_\textsuperscript{D} f_\textsuperscript{S}(1-f_\textsuperscript{D})$, the integrand $I(\omega,\omega')$ in Eq.~(\ref{eqcot}) is only non vanishing at low temperatures in the blue region of the $\omega-\omega'$ plane, depicted in Fig.~\ref{fig:cotunnel}(c). Upon increasing temperature, the product is also non-vanishing along the orange stripes and on the orange spot.

In Fig.~\ref{fig:cotunnel}(c) the roots of the denominators are represented by dashed lines. It is evident that the integral of $\hat K_\textsuperscript{EC}$ has a large magnitude only in the case the root line $\omega=\omega'$ and the colored regions meet when varying the bias voltage. Thus at low temperatures and $V_\textsuperscript{SD}=0$ no transport is possible as the corner of the blue region and the $\omega=\omega'$ line cannot touch. Upon increasing temperature, transport is accessible through the orange regions at $\omega'=\mu_\textsuperscript{D}-\Delta$ and $\omega=\mu_\textsuperscript{S}-\Delta$, see scheme in Fig.~\ref{fig:cotunnel}(c). This corresponds to the gate independent resonance at zero bias. In this simple resonance picture we obtain the elastic forward cotunneling rate (see appendix E) in the middle of a Coulomb diamond by a first approximation of the integrand in Eq.~(\ref{eqcot})
\begin{eqnarray}
 \Gamma_\textsuperscript{EC}^{\chi\to\chi}&=&N_\chi\hbar\left(\frac{2}{U}\right)^2 \Gamma_S \Gamma_D \int \frac{d\omega}{2\pi}  D(\omega,\Delta)D(\omega+eV_\textsuperscript{SD},\Delta) \nonumber \\* 
&&\times f(\omega)[1-f(\omega+eV_\textsuperscript{SD})],\label{cotton}
\end{eqnarray}
where we directly pointed out the bias dependence of the rate and introduced a degeneracy factor $N_\chi$ depending on the state $|\chi\rangle$. Also including the backward process the linear conductance is then approximated by
\begin{eqnarray}
 G=\frac{dI}{dV_\textsuperscript{SD}}\Bigg|_{V_\textsuperscript{SD}=0}\approx N_\chi\frac{e^2}{\hbar}\left(\frac{2}{U}\right)^2 \frac{\hbar^2\Gamma_S \Gamma_D}{k_BT} \int \frac{d\omega}{2\pi}  D^2(\omega,\Delta)f(\omega)f(-\omega).
\end{eqnarray}
This expression already shows a Boltzmann like behavior $\exp[-\Delta/(k_BT)]$ for low temperatures $T\ll\Delta/k_B$ and reproduces the normal conducting result $G=N_\chi\frac{e^2}{h}\frac{\hbar^2\Gamma_S\Gamma_D}{U^2}$ in the limit $\Delta\ll k_BT$. In particular, the former asymptotic characteristics indicates a transport property based on thermal excitation.

Analogously, subgap thermal replicas of the standard inelastic cotunneling lines are expected. We present a detailed analysis of the inelastic processes in appendix F and quote here the approximate result for the inelastic cotunneling rate
\begin{eqnarray}
 \Gamma_\textsuperscript{EC}^{\chi\to\chi'}&=&N_\chi\hbar\left(\frac{2}{U}\right)^2 \Gamma_S \Gamma_D \int \frac{d\omega}{2\pi}  D(\omega,\Delta)D(\omega-\delta+eV_\textsuperscript{SD},\Delta) \nonumber \\* 
&&\times f(\omega)[1-f(\omega-\delta+eV_\textsuperscript{SD})],
\end{eqnarray}
similar to what was found in Ref.~\cite{grove09}.

\section{Comparison of theoretical and experimental predictions}
\begin{figure}[tb]
\centering
\includegraphics[width=\columnwidth]{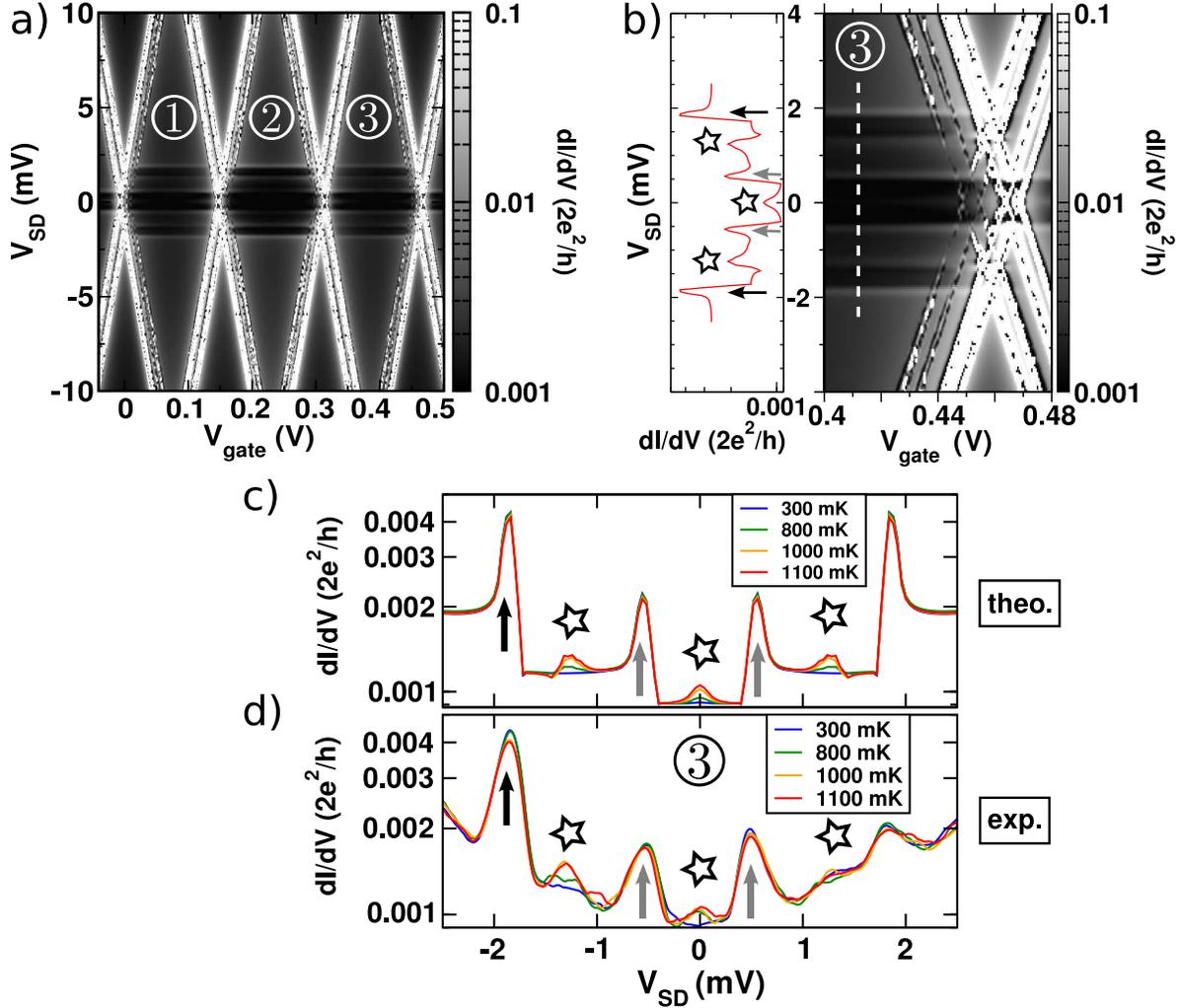}
\caption{(a) Calculated differential conductance of a CNT with level splitting $\delta=1.3\,$meV and charging energy $E_\textsuperscript{C}=15\,$meV. The temperature is $T=1.7\,$K and the BCS gap $\Delta=0.26\,$meV. The onset of inelastic and elastic cotunneling at $V_\textsuperscript{SD}=\pm(2\Delta+\delta)/e$ and $V_\textsuperscript{SD}=\pm2\Delta$, respectively, yields horizontal transition lines. Also gate independent features at bias voltages $V_\textsuperscript{SD}=\pm\delta/e$ and at zero bias can be pointed out. (b) Right panel: Zoom into the right corner of diamond \textcircled{\bf\scriptsize{3}} indicated in (a). Left panel: Bias trace corresponding to the gate voltage marked by the dashed white line in the right panel. In the bias trace the peaks indicated by stars are due to thermally activated quasiparticles. (c) Calculated bias traces for different temperatures. The peaks marked by stars correspond to thermal replicas of the standard cotunneling processes. To compare with the experiment we add 
a conductance offset 
of about $0.002\,e^2/h$ to our numerical data. (d) Equivalent experimental data for comparison. The bias-dependent background results from the gradual increase of the conductance in vicinity of the diamond edges.}\label{fig:stab}
\end{figure}

In the following we use the BCS gap $\Delta$, the excitation energy $\delta$, and the charging energy $E_\textsuperscript{C}$ extracted from the measured differential conductance plots to calculate the current through the CNT by means of the generalized master equation. Since the measured data revealed a relatively large critical temperature we could assume a temperature independent gap size in the considered temperature regime $T<T_c/2$. The calculations are performed by approximating the divergent DOS $D_l(\omega, \Delta)$ with a smoothened function~\bibnote{We replace the Heaviside function $\Theta(|\omega|-\Delta)\to\frac{1}{\exp(\gamma^{-1}(\omega+\Delta))+1}+\frac{1}{\exp(\gamma^{-1}(-\omega+\Delta))+1}$ by a blurred step function. Despite $\gamma$ is introduced empirically in this work, it can be shown that higher order processes involving quasiparticles lead to level broadening in the quantum dot and thus also to regularization of the divergence caused by the BCS density of states~\cite{levy97} similar to that provided by $\gamma$ here.} controlled by an empirical parameter $\gamma$ similar to the Dynes parameter~\cite{Dynes78}. A good fit to the experimental data for sample \textsf{A} is obtained by $\gamma\approx5.0\,\mu$eV, a coupling strength $\hbar\Gamma=0.01\,$meV and a conversion factor $\alpha=0.1$ for the gate voltage. The results of our transport calculations for sample \textsf{A} are shown in Fig.~\ref{fig:stab}(a)-(c) for temperature $T=1.7\,$K, such that $k_BT/\Delta=0.56$. 
Fig.~\ref{fig:stab}(d) shows the corresponding experimental data for diamond \textcircled{\bf\footnotesize{3}}. A short analysis of diamond \textcircled{\bf\footnotesize{2}} is given in the appendix B.
In the bias and gate voltage range of Fig.~\ref{fig:stab}(a) pronounced sequential tunneling lines and elastic and inelastic cotunneling features are seen. For a better resolution we restrict the gray scale of the differential conductance below the maximum value. In Fig.~\ref{fig:stab}(b) we focus on the Coulomb diamond denoted \textcircled{\bf\footnotesize{3}}. Beside the density plot we show the bias trace taken at the gate voltage marked by a white line, which supports the good quantitative agreement with the experimental data of Fig.~\ref{fig:sampleA}(d). The standard cotunneling peaks (arrows) as well as their thermal replicas (stars) can be clearly recognized. The thermal behavior of the cotunneling features is illustrated in Figs.~\ref{fig:stab}(c),(d) where the calculated and the measured differential conductance curves for different temperatures are presented. For the calculated curves we choose the same gate voltage as for the white dashed line in Fig.~\ref{fig:stab}. For the experimental data we 
averaged over a series 
of gate voltages marked by the box in Fig.~\ref{fig:sampleA}(d). In both cases we emphasize that the standard cotunneling peaks are almost temperature independent, whereas the thermal replicas at zero bias and at $V_\textsuperscript{SD}=\pm\delta/e$ rise with increasing temperature.

\section{Conclusions}
In summary, we report on new cotunneling transport properties of a CNT contacted to two superconducting Nb leads based on thermally assisted quasiparticle tunneling. We observe the thermal replica of the elastic and inelastic cotunneling resonances with increasing temperature above $600\,$mK. These lead to an extra zero-bias peak and to an inelastic peak corresponding to the lowest excitation energy in the $dI/dV$ characteristics. To explain these non-equilibrium phenomena we derive a generalized master equation based on the RDM approach in the charge conserved regime, applicable to any intradot interaction and finite superconducting gap. Modeling the CNT with a low-energy interacting spectrum, we find a remarkable agreement with the experimental results concerning the thermal behavior of the additional cotunneling peaks.

\ackn

 \input acknowledgement.tex


\appendix

\section{Experimental data of sample \textsf{B}}
\begin{figure}[tb]
\centering
\includegraphics[width=\columnwidth]{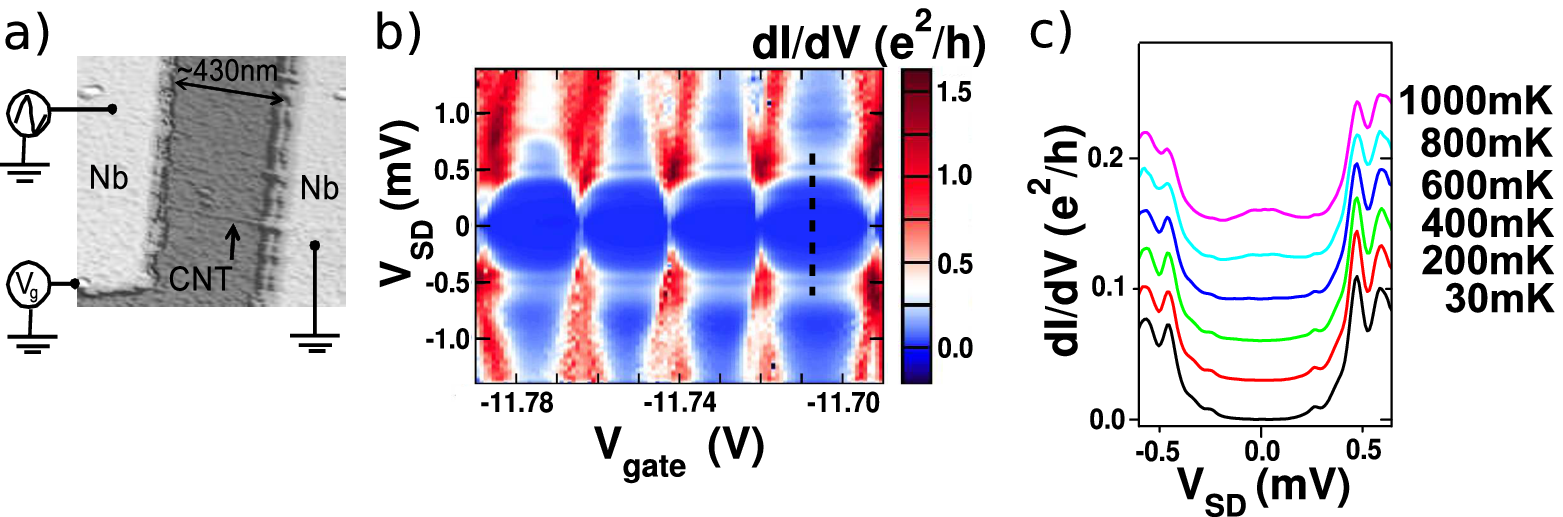}
\caption{(a) Atomic force micrograph of device \textsf{B}. (b) Differential conductance at $T=30\,$mK as function of bias voltage and back gate voltage. (c) Differential conductance curves in the low bias regime for different temperatures at gate voltage $V_\textsuperscript{gate}\approx -11.71\,$V (black dashed line in b)). The bias trace shows a zero-bias peak emerging at increasing temperature. To see the feature more clearly a conductance offset of about $0.03\,e^2/h$ was added systematically to each curve. The zero bias peak is accompanied by the elastic and inelastic cotunneling peaks at negative and positive bias.}\label{fig:amit}
\end{figure}

We have in addition confirmed the prediction of a zero bias peak due to thermally excited elastic cotunneling in an other experimental setup. The description of sample \textsf{B} can be found in the main text. An atomic force micrograph of the studied quantum dot device is shown in Fig.~\ref{fig:amit}(a).

We observe regular Coulomb blockade diamonds over a large gate voltage range, also suggesting a defect free CNT. In Fig.~\ref{fig:amit}(b) we show high resolution measurements for a selected gate range including four Coulomb diamonds at temperature $T=30\,$mK. Inside the Coulomb diamonds we can identify gate independent transition lines suggesting a symmetric coupling to the superconducting leads. To clarify the bias threshold of these horizontal lines, we take a bias trace of the interesting region at a fixed gate voltage $V_\textsuperscript{gate}\approx -11.71\,$V pointed out by the dashed line. This enables us to observe the onset of a stable conductance peak for temperatures above $T\approx600\,$mK which is more and more pronounced with increasing temperature. For that reason we assign the gate independent conductance peak to a thermally assisted elastic cotunneling process. A detailed theoretical discussion follows in D. However, a thermal replica of the inelastic cotunneling peak at bias voltage $V_\textsuperscript{SD}=\pm 0.11\,$meV cannot be clearly seen. This may be due to an overlap with the zero bias peak.

\section{Analysis of the Coulomb diamond \textcircled{\bf\footnotesize{2}} of sample \textsf{A}}
\begin{figure}[tb]
\centering
\includegraphics[width=.7\columnwidth]{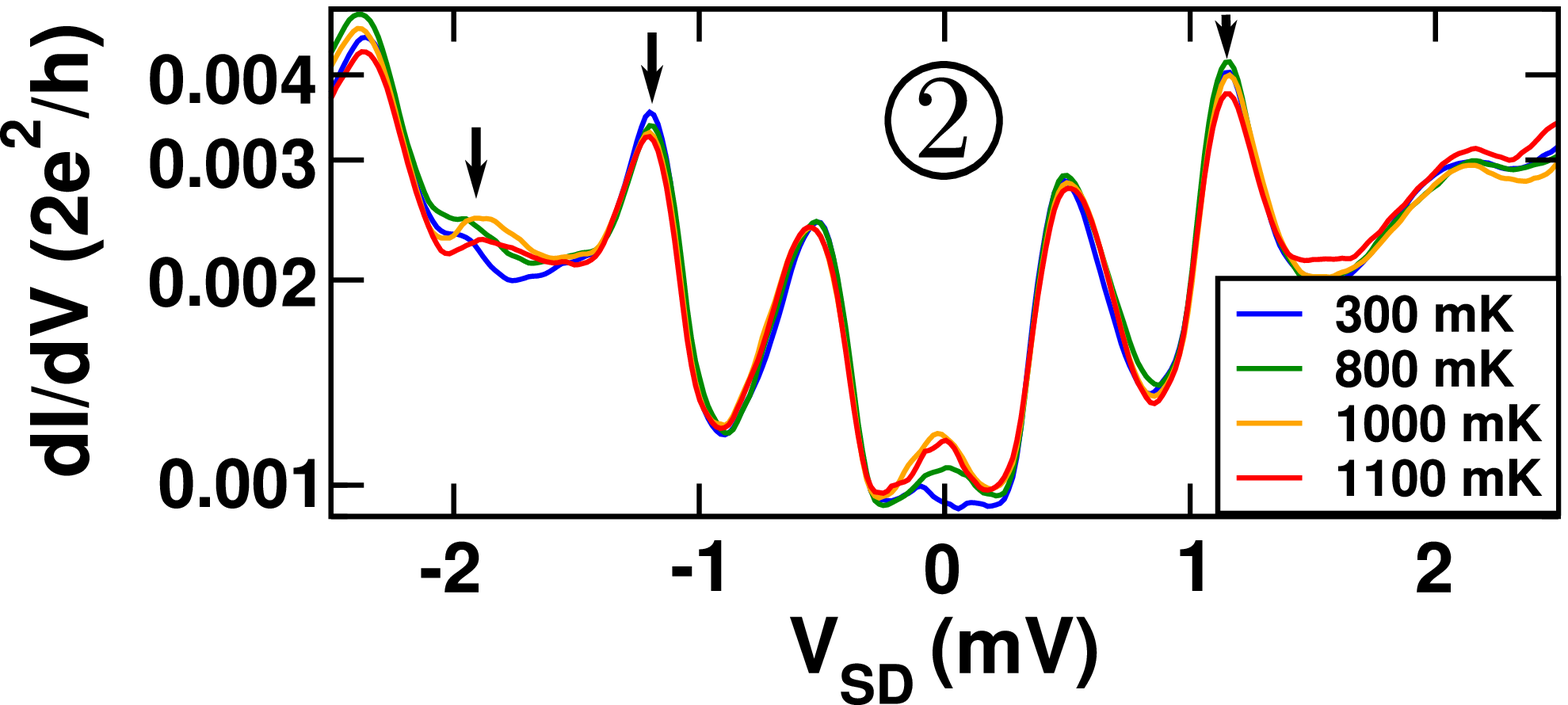}
\caption{Measurement of the differential conductance of sample \textsf{A} taken in the middle of diamond \textcircled{\bf\scriptsize{2}} for different temperatures. More features are observed than for diamond \textcircled{\bf\scriptsize{3}} indicating a more complex excitation structure of the CNT spectrum. We can see two additional excitation transition resonances, indicated by arrows, not observed in diamond \textcircled{\bf\scriptsize{3}}.}\label{fig:2nd}
\end{figure}

In Fig.~\ref{fig:2nd} we show the bias trace of the measured differential conductance in the middle of the Coulomb diamond \textcircled{\bf\footnotesize{2}} defined in the main text. It was obtained by the same averaging procedure as for the bias trace of diamond \textcircled{\bf\footnotesize{3}} explained in the main text. The curves for different temperatures include a richer peak structure than for the other diamond. We can identify the standard elastic cotunneling peaks at $V_\textsuperscript{SD}=\pm 0.55\,$mV as well as the inelastic peaks at $V_\textsuperscript{SD}=\pm 2.2\,$mV. The shift of the bias threshold for the excitation energy $\delta$ in comparison to the diamond \textcircled{\bf\footnotesize{3}} can be explained by a gate-dependent spin-orbit coupling in multielectron carbon nanotubes \cite{jesp11}, as it can also be seen in the overview Figure 1b). Besides, we can clearly recognize the rise of the thermal elastic cotunneling peak at zero bias with increasing temperature.

The additional features at bias voltage $V_\textsuperscript{SD}=\pm1.2\,$mV cannot be explained by a single shell model. Since the peak height is not temperature dependent, it must be a standard inelastic cotunneling feature. For our calculations we thus have to include a more complex excitation spectrum where the splitting to the next higher shell is smaller than $\delta$. As it was shown in Ref. \cite{peck13}, an two particle ground state can lead to a rather complicated excitation spectrum where energetically close shells interact with each other resulting in an effective shell splitting smaller than $\delta$. Calculating such an effective Hamiltonian will remain a future task.

Also the small peak at bias voltage $V_\textsuperscript{SD}\approx- 1.9\,$mV is almost temperature independent. By inspection of the stability diagram, we classify it as a cotunneling assisted sequential tunneling process (COSET). In such a COSET an excited state is populated by a preceding inelastic cotunneling process yielding to a gate dependent sequential resonance peak inside the Coulomb blockade regime\cite{schl05,huttel09,Koller10}. A more detailed discussion is left to future work.

\section{BCS theory in the charge conserved regime}
In macroscopic superconductors with a large number of particles, the boson-like condensate is well described by a phase coherent state $|\Phi\rangle$ with definite phase $\Phi$. The presence of a relative phase between two weakly linked superconductors is at the origin of the Josephson effect~\cite{Jos62,ander63,jos74}. In mesoscopic superconductors, charging effects due to Coulomb interaction break the degeneracy of states with different number $M$ of Cooper pairs. In such cases the phase $\Phi$ becomes uncertain and one has to project the state $|\Phi\rangle$ onto a state $|2M\rangle$ with fixed Cooper pair number $M$~\cite{BarCooSch57,Sch99}. In this phase incoherent regime, the BCS Hamiltonian is thus properly diagonalized by means of the particle number conserving Bogoliubov-Valatin transformation
\begin{eqnarray} 
 \hat c^\dagger_{\vec k\sigma}&=&u_{\vec k}\hat\gamma^\dagger_{\vec k\sigma}+\sigma v^*_{\vec k}\hat S^\dagger\hat\gamma_{-\vec k\bar\sigma},\label{BVT1} \\*
 \hat c_{\vec k\sigma}&=&u^*_{\vec k}\hat\gamma_{\vec k\sigma}+\sigma v_{\vec k}\hat S\hat\gamma^\dagger_{-\vec k\bar\sigma},\label{BVT2}
\end{eqnarray}
including quasiparticle, $\hat \gamma^{(\dagger)}_{\vec k\sigma}$, as well as Cooper pair, $\hat S^{(\dagger)}$, operators. From the fermionic excitations described by the quasiparticle operators we demand $\{\hat\gamma_{\vec k\sigma},\hat\gamma^\dagger_{\vec k'\sigma'}\} =\delta_{\vec k \vec k'}\delta_{\sigma\sigma'}$. Moreover, the Cooper pair condensate and the quasiparticles are decoupled, i.e.
\begin{eqnarray}
 \left[\hat S^{(\dagger)},\hat\gamma^{(\dagger)}_{\vec k'\sigma'}\right]&=&0.\label{commutator}
\end{eqnarray}
By means of these commutator relations, we can further show that for the number operator $\hat N$ of the electrons it holds
\begin{eqnarray}
 \left[\hat N,\hat S^\dagger\right]&=&2\hat S^\dagger,
\end{eqnarray}
i.e., the Cooper pair operator keeps the system in a state with a well defined charge number:
\begin{eqnarray}
 \hat S\left|2M\right\rangle &=& \left| 2M-2\right\rangle.\label{eq:CP}
\end{eqnarray}
Together with Eq.~(\ref{commutator}), we conclude that the Cooper pair condensate is the vacuum state for the quasiparticles and that fermionic excitations can be described by
\begin{eqnarray}
 \hat\gamma^\dagger_{\vec k\sigma} \left|0,2M\right\rangle &=&\left|\vec k\sigma,2M\right\rangle, \\*
 \hat\gamma_{\vec k\sigma} \left|0,2M\right\rangle &=&0.\label{eq:QP}
\end{eqnarray}
In the phase incoherent regime, the equilibrium grand canonical density operator of the superconductor is given by
\begin{eqnarray}
 \hat\rho_R=\frac{e^{-\beta\hat H_\textsuperscript{gc}}}{Z},\label{eq:RDO}
\end{eqnarray}
with $\beta^{-1}\equiv k_\textsuperscript{B}T$ the inverse temperature. Here $Z\equiv\textnormal{Tr}_\textsuperscript{R}\left(e^{-\beta\hat H_\textsuperscript{gc}}\right)$ is the partition function, where we introduced the grand canonical Hamiltonian $\hat H_\textsuperscript{gc}\equiv\hat H-\mu \hat N$, and $\hat H$ is as defined in Eq.~(2) of the main text. Accounting for the properties  Eqs.~(\ref{eq:CP})-(\ref{eq:QP}) of the quasiparticle and Cooper pair operators, the calculation of the thermal expectation value
\begin{eqnarray}
 \left\langle\hat O\right\rangle&\equiv&\Tr_\textsuperscript{R}\left(\hat\rho_\textsuperscript{R}\hat O\right)  \nonumber\\
  &=&\sum_{\{n_{\vec k\sigma}\},M}\left\langle\{n_{\vec k\sigma}\},2M\right|\hat\rho_\textsuperscript{R}\hat O\left|\{n_{\vec k\sigma}\},2M\right\rangle,
\end{eqnarray}
of an operator $\hat O$ in the basis $\{|\{n_{\vec k\sigma}\},2M\rangle\}$ of the superconducting lead remains a standard task.

In the main text we claimed that the superconducting gap is not vanishing in the charge conserved regime. The statement can be proved in the following way:
\begin{eqnarray}
 \Delta&\equiv&|V|\sum_{\vec k}\left\langle \hat S^\dagger\hat c_{-\vec k\downarrow}\hat c_{\vec k \uparrow}\right\rangle \nonumber \\*
&=&|V|\sum_{\vec k}\Tr_\textsuperscript{R}\left(\hat\rho_\textsuperscript{R}\hat S^\dagger\left(u^*_{-\vec k}u^*_{\vec k}\hat\gamma_{-\vec k\downarrow}\hat\gamma_{\vec k \uparrow}-v_{-\vec k}v_{\vec k}\hat S\hat\gamma^\dagger_{\vec k\uparrow}\hat S\hat\gamma^\dagger_{-\vec k\downarrow}-u^*_{\vec k}v_{-\vec k}\hat S\hat\gamma^\dagger_{\vec k\uparrow}\hat\gamma_{\vec k\uparrow}\right.\right. \nonumber \\*
&&\left.\left.+u^*_{-\vec k}v_{\vec k}\hat\gamma_{-\vec k\downarrow}\hat S\hat\gamma^\dagger_{-\vec k\downarrow}\right)\right) \nonumber \\*
&=&|V|\sum_{\vec k} -u^*_{\vec k}v_{-\vec k}f(E_{\vec k})+u^*_{-\vec k}v_{\vec k}\left(1-f(E_{-\vec k})\right),\hspace{1em}
\end{eqnarray}
where in the last line we have used the orthogonality $\langle2M|(\hat S^\dagger\hat S)|2M\rangle=(1-\delta_{M0})$ of the Cooper pair states, and $\langle\vec k\sigma|\vec k'\sigma'\rangle=\delta_{\vec k\vec k'}\delta_{\sigma\sigma'}$ of the quasiparticle states. Moreover, $\langle\hat\gamma^\dagger_{\vec k\uparrow}\hat\gamma_{\vec k\uparrow}\rangle=f(E_{\vec k})$, with $f(x)=1/(e^{x}+1)$. Indeed the superconducting gap has a finite magnitude whose value depends on the temperature, as known from the BCS theory.

In the same manner it can be shown that the expectation values $\langle\hat c^\dagger_{\vec k\sigma}\hat c^\dagger_{\vec k'\sigma'}\rangle$ and $\langle\hat c_{\vec k\sigma}\hat c_{\vec k'\sigma'}\rangle$ vanish in the charge conserved regime.

\section{Transport in the charge conserved regime}
For superconducting leads in the phase incoherent regime charge is conserved. This fact has important consequences when looking at quantum transport through a quantum dot coupled to such charge conserved BCS leads. Because the tunneling Hamiltonian $\hat H_T$ (Eq.~(6) of the main text) and the quantum dot Hamiltonian are also charge conserving, charge is conserved during transport. As a consequence, the quantum dot density operator $\hat\rho_\textsuperscript{red}$ is block diagonal in the charge representation. In other words, there are no coherences between states with different numbers of Cooper pairs. Let us emphasize that Cooper pairs still take part in tunneling events, as we will show in the next section when analyzing one contribution to elastic cotunneling in the transport characteristics.

\section{Analysis of the elastic cotunneling diagram}
From a standard evaluation of the multiple commutators~\cite{Koller10} constituting the kernel $\hat K(t,\tau)$, we obtain for the matrix element $(\hat K_\textsuperscript{EC})^{\chi\chi}_{\chi\chi}\equiv \int^\infty_0 dt'\,e^{-0^+t'}\langle\chi|\hat K(t,t-t')[|\chi\rangle\langle\chi|]|\chi\rangle$ the expression
\begin{figure}[tb]
\centering
\includegraphics[width=.4\columnwidth]{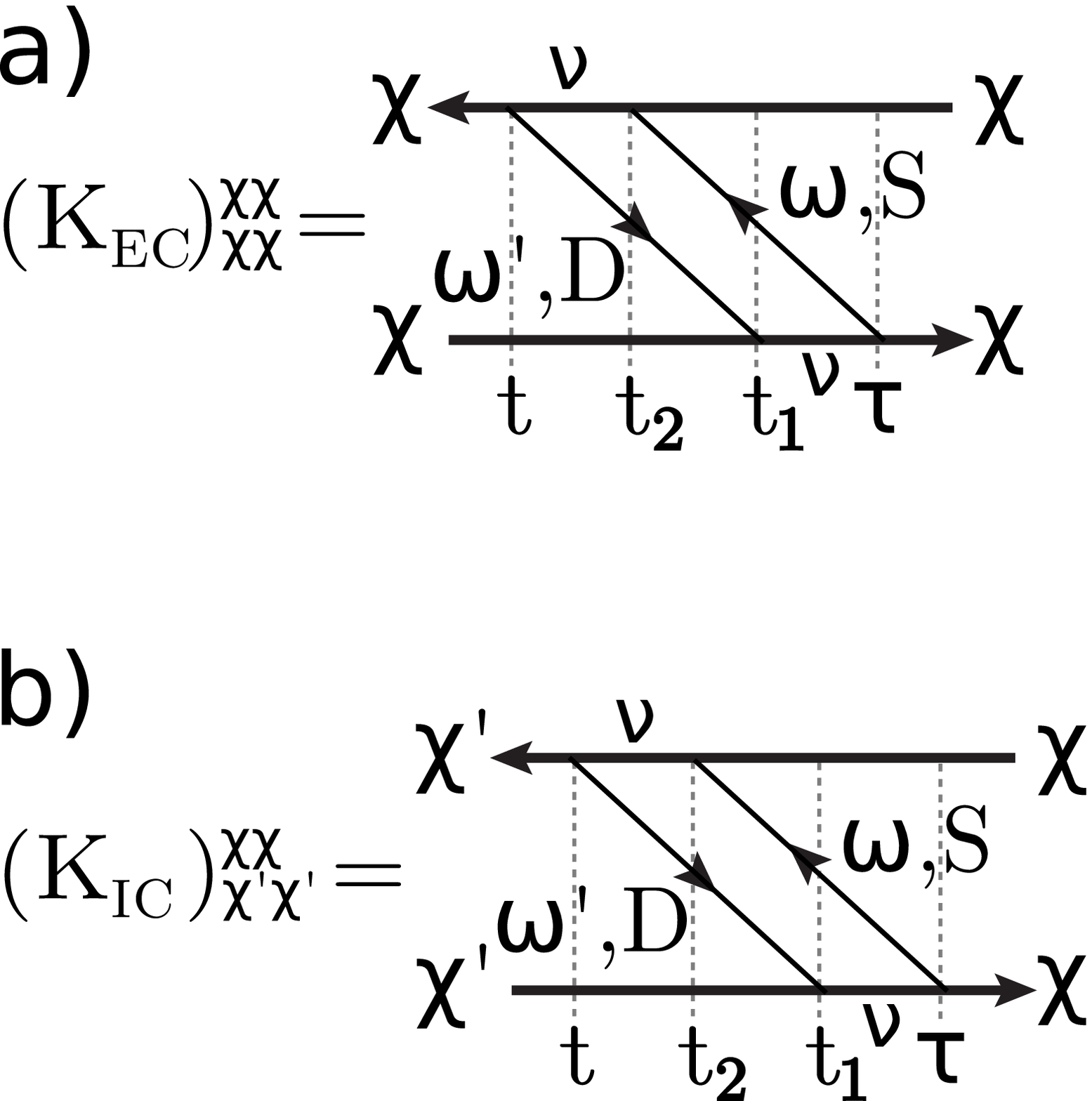}
\caption{Diagrammatic representation of one relevant contribution to elastic (a) and inelastic (b) cotunneling. Necessary for the inelastic part is the energetic excitation of the final state $\chi'$ in comparison to the initial state $\chi$, with both states having the same amount of charge.}\label{fig:recot}
\end{figure}
\begin{eqnarray}
 \left(\hat K_\textsuperscript{EC}\right)^{\chi\chi}_{\chi\chi}
&=&\sum_{\vec k\sigma,\vec k'\sigma'}\sum_{mm'\nu}\int^\infty_0 dt'\,e^{-0^+t'}\int^{t'}_0 dt'_1 \int^{t'_1}_0 dt'_2 \left\langle\hat c^\dagger_{S\vec k\sigma}(\tau)\hat c_{S\vec k\sigma}(t_2)\right \rangle \nonumber \\*
&&\left\langle\hat c_{D\vec k'\sigma'}(t_1)\hat c^\dagger_{D\vec k'\sigma'}(t)\right \rangle \frac{|T_S|^2|T_D|^2}{\hbar^4}\nonumber \\*
&&\times\langle\chi|\hat d_{m'\sigma'}(t)|\nu\rangle\langle\nu|\hat d^\dagger_{m\sigma}(t_2)|\chi\rangle\langle\chi|\hat d_{m\sigma}(\tau)|\nu\rangle \langle\nu|\hat d^\dagger_{m'\sigma'}(t_1)|\chi\rangle, \nonumber
\end{eqnarray}
whose diagrammatic representation is shown in Fig.~\ref{fig:recot}(a). For the time differences we used the notation $t'\equiv t-\tau$, $t'_1\equiv t-t_1$ and $t'_2\equiv t-t_2$. When evaluating the expectation values of the electron lead operators we need the Bogoliubov transform, Eqs.~(\ref{BVT1}) and (\ref{BVT2}). We then see that also Cooper pairs give contributions to the transport process through the acting operators $\hat S^{(\dagger)}$. We find:
\begin{eqnarray}
 &&\left\langle\hat c^\dagger_{S\vec k\sigma}(\tau)\hat c_{S\vec k\sigma}(t_2)\right \rangle \nonumber \\*
 &=& \left\langle\left(u_{S\vec k}\hat\gamma^\dagger_{S\vec k\sigma}(\tau)+\sigma v^*_{S\vec k}\hat S^\dagger_S(\tau)\hat\gamma_{S-\vec k\bar\sigma}(\tau)\right)\right. \nonumber \\*
&&\left. \times\left(u^*_{S\vec k}\hat\gamma_{S\vec k\sigma}(t_2)+\sigma v_{S\vec k}\hat S_S(t_2)\hat\gamma^\dagger_{S-\vec k\bar\sigma}(t_2)\right)\right\rangle \nonumber \\*
&=&|u_{S\vec k}|^2 \,\exp[-\frac{i}{\hbar}(E_{S\vec k}+\mu_S)(t'-t'_2)]\langle\hat\gamma^\dagger_{S\vec k\sigma}\hat\gamma_{S\vec k\sigma}\rangle \nonumber \\*
&&+|v_{S\vec k}|^2 \,\exp[\frac{i}{\hbar}(-E_{S\vec k}-\mu_S)(t'-t'_2)]\langle\hat S^\dagger_S\hat\gamma_{S-\vec k\bar\sigma}\hat S_S\hat\gamma^\dagger_{S-\vec k\bar\sigma}\rangle \nonumber \\*
&=&|u_{S\vec k}|^2 \,\exp[-\frac{i}{\hbar}(E_{S\vec k}+\mu_S)(t'-t'_2)][\exp(\beta E_{S\vec k})+1]^{-1} \nonumber \\*
&&+|v_{S\vec k}|^2 \,\exp[-\frac{i}{\hbar}(E_{S\vec k}+\mu_S)(t'-t'_2)][\exp(\beta E_{S\vec k})+1]^{-1}  \nonumber \\*
&=& \exp[-\frac{i}{\hbar}(E_{S\vec k}+\mu_S)(t'-t'_2)][\exp(\beta E_{S\vec k})+1]^{-1}, \nonumber
\end{eqnarray}
where in the last line we used the normalisation condition $|u_{l\vec k}|^2+|v_{l\vec k}|^2 =1$. The second expectation value can be calculated in a similar way such that for the kernel component of the elastic cotunneling we obtain
\begin{eqnarray}
 \left(\hat K_\textsuperscript{EC}\right)^{\chi\chi}_{\chi\chi}&\equiv&\sum_{\vec k\vec k'}\sum_\nu\int^\infty_0 dt'_2\int_{t'_2}^\infty dt'_1 \int_{t'_1}^\infty dt'\,e^{-0^+t'}  \nonumber \\*
&&\times\exp[-\frac{i}{\hbar}(E_{S\vec k}+\mu_S)(t'-t'_2)]\, \exp[\frac{i}{\hbar}(E_{D\vec k'}+\mu_D)t'_1]  \nonumber \\*
&&\times (\exp[\beta E_{S\vec k}]+1)^{-1}(\exp[-\beta E_{D\vec k'}]+1)^{-1}\exp[\frac{i}{\hbar}(E_\chi-E_\nu) t'_2] \nonumber \\*
&& \times\exp[-\frac{i}{\hbar}(E_\chi-E_\nu) (t'-t'_1)] \frac{|T_S(\chi,\nu)|^2|T_D(\chi,\nu)|^2}{\hbar^4}\nonumber \\*
&=&-i\hbar\Gamma_S\Gamma_D \sum_\nu\int \frac{d\omega}{2\pi}\frac{d\omega'}{2\pi}D_S(\omega,\Delta)D_D(\omega',\Delta) \nonumber \\*
&&\times \frac{f_S(\omega)(1-f_D(\omega'))}{(-\omega+E_\nu-E_\chi+i0^+)(\omega'-\omega+i0^+)(\omega'+E_\chi-E_\nu+i0^+)} \nonumber \\*
&\equiv& -i\hbar\Gamma_S\Gamma_D \sum_\nu\int \frac{d\omega}{2\pi}\frac{d\omega'}{2\pi}\, I(\omega,\omega'), \label{supeqcot} 
\end{eqnarray}
which is the same result as in the main text with $T_l(\chi,\nu)\equiv\sqrt{\tilde\rho_l}\sum_{m\sigma}T_l\langle\chi|\hat d_{m\sigma}|\nu\rangle$, and the electron density of states $\tilde\rho_l$ in lead $l$. In the last step the variable transformation $\tilde t_1\equiv t'_1-t'_2$, $\tilde t\equiv t'-t'_1$ was applied in order to decouple the three time integrations. Besides, we expressed the energies $\omega$, $\omega'$ with respect to the electrochemical potential $\mu_l$.

\begin{figure}[tb]
\centering
\includegraphics[width=.8\columnwidth]{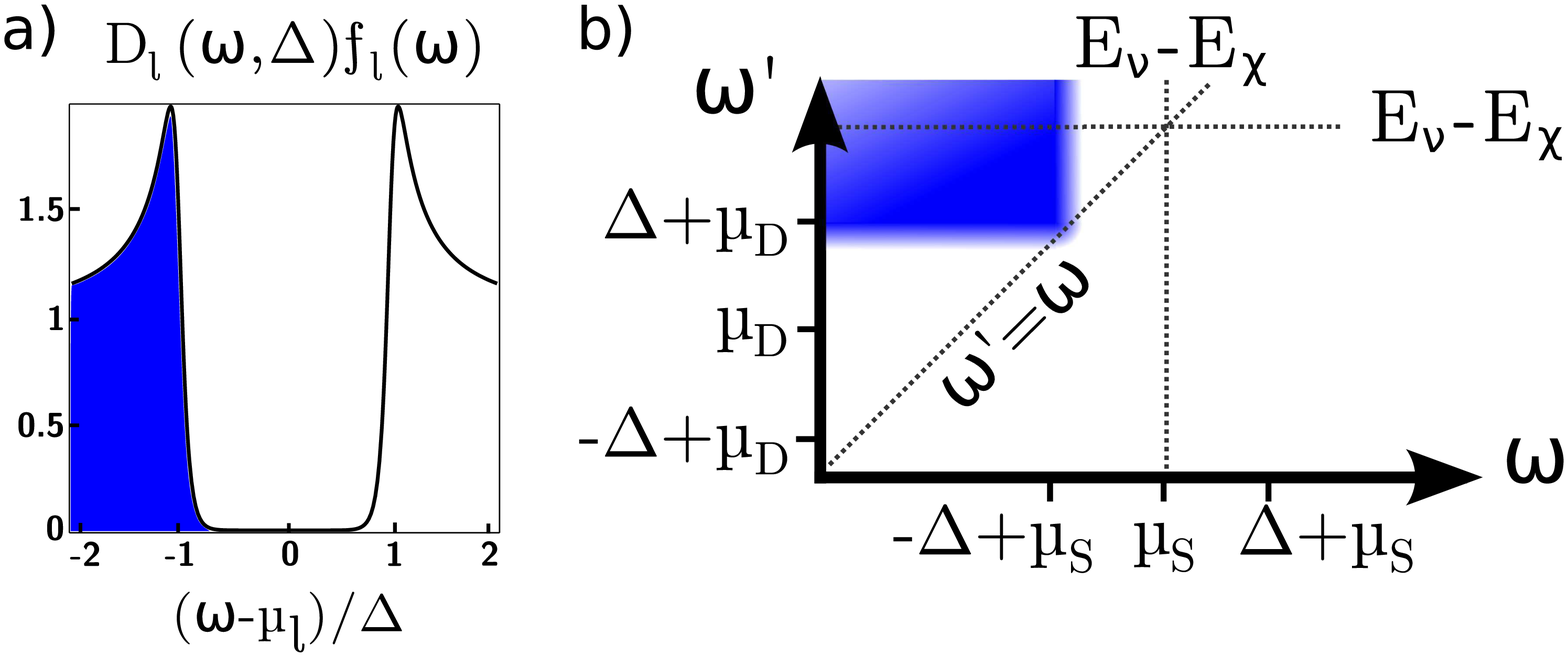}
\caption{(a) Product of the Fermi function and the BCS density of states for low temperatures. (b) Integrand $I(\omega,\omega')$ occurring in the two dimensional integral of Eq.~(\ref{supeqcot}). The three dashed lines correspond to the roots of the denominator of $I(\omega,\omega')$. The figure shows the parameter regime at low temperatures $T\ll\Delta/k_B$ and finite bias $V_\textsuperscript{SD}=(\mu_S-\mu_D)/e>0$ at which the onset of elastic cotunneling occurs. When the bias voltage is set such that $-\Delta+\mu_S=\Delta+\mu_D$, the corner of the blue region meets the root line $\omega=\omega'$, as shown in the figure, yielding the threshold for elastic cotunneling.}\label{fig:condecot1}
\end{figure}

\begin{figure}[tb]
\centering
\includegraphics[width=.8\columnwidth]{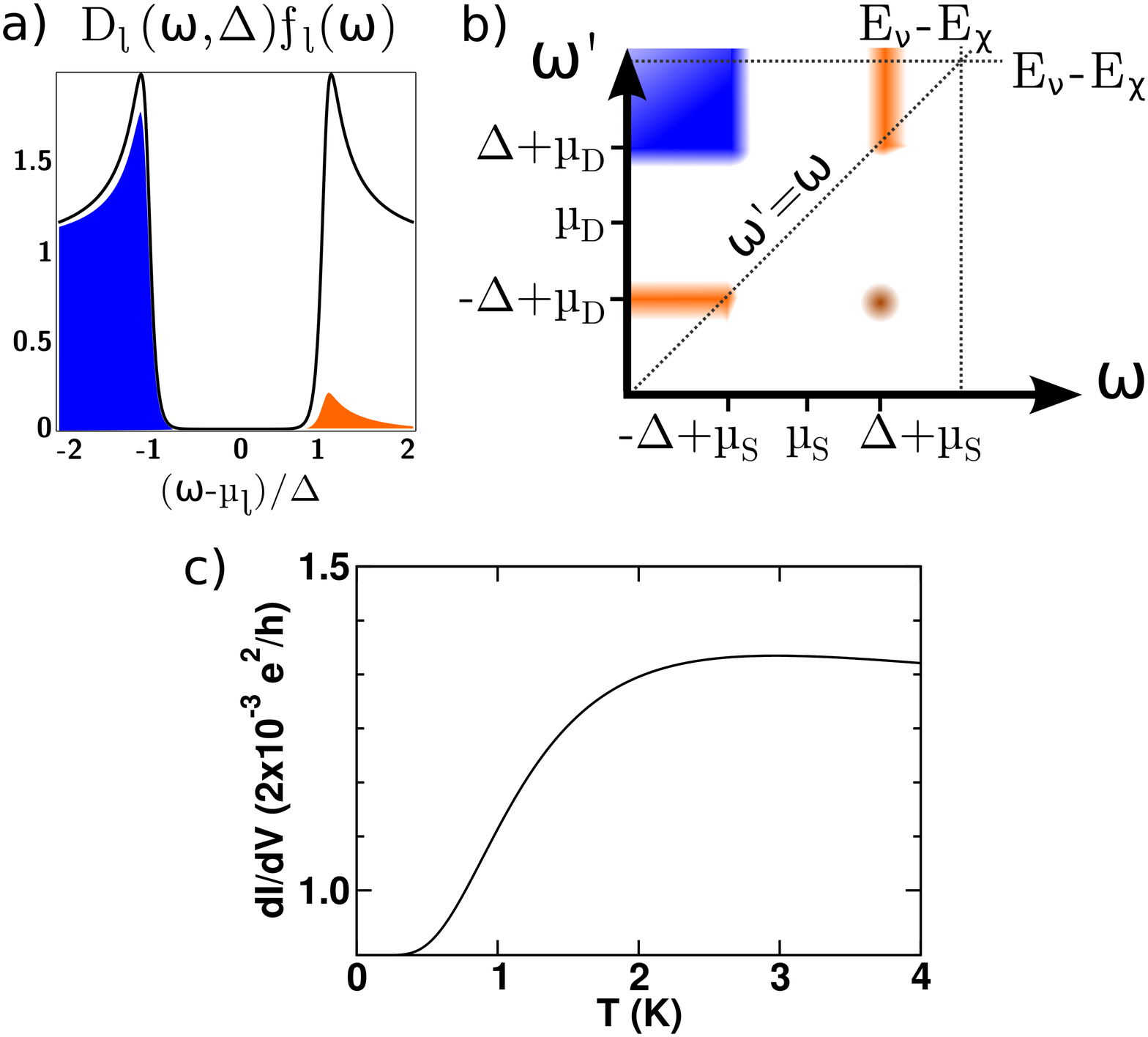}
\caption{(a) Product of the Fermi function and the BCS density of states for high temperatures. (b) The figure shows the parameter regime at high temperatures $T\lesssim\Delta/k_B$ and $V_\textsuperscript{SD}=0$ at which thermally assisted elastic cotunneling occurs. When the root line $\omega=\omega'$ hits the corners of the orange area, it holds $-\Delta+\mu_S=-\Delta+\mu_D$, corresponding to zero bias. Thus, even in the zero bias regime, thermally excited elastic cotunneling features emerge which are absent for low temperatures. The dark orange color is only important for high temperatures $T\gtrsim\Delta/k_B$. (c) Temperature dependence of the appearing zero bias peak in the stability diagram. For comparison we add a conductance offset of about 0.002 $e^2/h$ to our numerical data. For small temperatures $T\ll\Delta/k_B$ a Boltzmann like behavior $\exp[-\Delta/(k_BT)]$ can be identified. }\label{fig:condecot2}
\end{figure}

To investigate the case when the double integral, and thus the kernel component, gives a relevant contribution to the transport dynamics, we analyze the integrand in detail. We are mainly interested in the region in the bias and gate voltage range in which the system is blocked into the ground state of the corresponding Coulomb blockade region in the sense of the sequential tunneling limit. If $N$ electrons are trapped in the $N$-Coulomb diamond, the condition for strong Coulomb blockade is $\mu_{S/D}-\Delta\ll E_{N\pm1}-E_N$. In our example the ground state energy for $N$ charges is $E_\chi$, while $E_\nu$ is the energy of the ($N+1$)-particles state $|\nu\rangle$. Hence, in the blockade regime is $\mu_{S/D}-\Delta\ll E_{\nu}-E_\chi$. Moreover, taking the product of the Fermi function and the BCS density of states (Fig.~\ref{fig:condecot1}(a)) into account, only the blue region of the $\omega-\omega'$ plane, depicted in Fig.~\ref{fig:condecot1}(b), is relevant for the integrand $I(\omega,\omega')$ in Eq.~(\ref{supeqcot}) at low temperatures. Upon increasing temperature, the product of Fermi functions and BCS density of states in the integrand $I(\omega,\omega')$ is also non vanishing along the orange stripes (Figs.~\ref{fig:condecot2}(a) and (b)).
The colored regions in the figure are the relevant energy region where the product of the density of states and the Fermi functions $D_S D_D f_S(1-f_D)\gg 0$ is not vanishing.

In Figs.~\ref{fig:condecot1}(b) and \ref{fig:condecot2}(b) the roots of the denominators are represented by dashed lines. As explained in the main text, we are looking for the cases in which the roots meet the colored regions. In particular, the threshold for the onset of standard elastic cotunneling processes is obtained for those values of the bias voltage such that the $\omega=\omega'$ root touches the corner of the blue region (see Fig.~\ref{fig:condecot1}(b)). In that case the blue region includes the horizontal and the diagonal ($\omega'=\omega$) zeros of the denominators. Thus the bias threshold $V_\textsuperscript{SD}=\pm2\Delta/e$ for the low temperature regime is obtained, when the condition $\Delta+\mu_D=\omega'=\omega=-\Delta+\mu_S$ is used together with $\mu_S-\mu_D=eV_\textsuperscript{SD}$. Note that for this bias voltage the diagonal zeros are located at the corner of the blue region, as seen in Fig.~\ref{fig:condecot1}(b), where the product of density of state 
and of the Fermi function has its largest value resulting in a peak structure in the voltage characteristics.

For higher temperatures additional scenarios have to be taken into account as the orange regions in Fig.~\ref{fig:condecot2}(b) cannot be neglect anymore. Thus the condition for strong Coulomb blockade has to be adapted to the low bias regime, meaning $\mu_{S/D}+\Delta\ll E_{N\pm1}-E_N$, in order to prevent thermally excited sequential tunneling as shown in Ref. \cite{Gaass14}. For our case this yields $\mu_{S/D}+\Delta\ll E_\nu-E_\chi$. To see a rising of thermal elastic cotunneling the diagonal $\omega=\omega'$ root has to meet the orange region. Then one has to investigate the cases when the orange regions include the horizontal and the diagonal zeros of the denominators. In this situation we only need a minimal bias $|V_\textsuperscript{SD}|\ge 0$. Thus, for large enough temperatures a remarkable contribution of the component of the kernel $\hat K_\textsuperscript{EC}$ for the elastic cotunneling is always present in the bias-gate voltage range since the onset occurs at zero bias, as one can see by 
means of the condition 
$-\Delta+\mu_D=\omega'=\omega=-\Delta+\mu_S$.

After analyzing the property of the integrand of the kernel element we can give a first approximation for the elastic cotunneling rate in the middle of a Coulomb diamond for the process shown in Fig.~\ref{fig:recot}(a). As explained above we only have to investigate the integrand in the energy area, where $\omega\approx\omega'$. Furthermore we investigate the Coulomb diamond \textcircled{\bf\footnotesize{3}}, in particular, as in the main text, where the center is placed at a gate voltage $V_\textsuperscript{gate}=(\frac{5}{2}E_\textsuperscript{C}+\frac{1}{2}\delta)/e$. Then the energy difference $E_\nu-E_\chi=E_4-E^0_3=U/2$ with the ground state energy $E^0_3$ of the 3-particle state and $E_4$ of the 4-particle state. Thus in the energy region where the product of Fermi functions $f_S(\omega)[1-f_D(\omega)]$ is non-zero the denominator is almost constant with magnitudes $\pm U/2$. This leads to the result
\begin{eqnarray}
 \Gamma_\textsuperscript{EC}^{3\to3}&\approx&-\hbar\int \frac{d\omega}{2\pi}D_S(\omega,\Delta)D_D(\omega,\Delta)\frac{\Gamma_S\Gamma_Df_S(\omega)(1-f_D(\omega))}{(-\omega+E_4-E^0_3)(\omega+E^0_3-E_4)} \nonumber \\*
&\approx& \hbar\left(\frac{2}{U}\right)^2\Gamma_S\Gamma_D\int \frac{d\omega}{2\pi}D_S(\omega,\Delta)D_D(\omega,\Delta)f_S(\omega)[1-f_D(\omega)].\nonumber
\end{eqnarray}
In a last step one could transform the parameter $\omega$ to obtain the bias voltage $V_\textsuperscript{SD}=(\mu_S-\mu_D)/e$ as in Eq.~(\ref{cotton}). The result of a calculation of the linear cotunneling conductance in terms of the rate expression above at zero bias is shown in Fig.~\ref{fig:condecot2}(c).

There are other diagrams contributing to elastic cotunneling. However, using the diagrammatic rules to evaluate their analytic expression, one realizes that they contain two different intermediate states with one unit of charge more and less than the state $|\chi\rangle$; hence the two zeros of the corresponding denominator in the integrand are energetically far away from each other, resulting in a smaller contribution to the integral.

We also wish to mention the dark orange dot in Fig.~\ref{fig:condecot2}(b). In that region the kernel component $\hat K_\textsuperscript{EC}$ contributes at high temperatures only. This can be explained for the case the diagonal zeros hit the area resulting in a condition $-\Delta+\mu_D=\omega'=\omega=\Delta+\mu_S$ for the bias threshold $V_\textsuperscript{SD}=-2\Delta/e$. The conductance peak in this bias region corresponds to an onset of a resonant charge current based on thermally excited quasiparticles in the drain producing unoccupied states in the low energy branch of the BCS density of states for even large temperatures.

\section{Inelastic cotunneling contributions}
In the same manner as in the section before, we can investigate leading contributions to the inelastic cotunneling. To this extent we identify the diagram shown in Fig.~\ref{fig:recot}(b) as one relevant inelastic cotunneling contribution to the kernel component $(\hat K_\textsuperscript{IC})^{\chi\chi}_{\chi'\chi'}$. Here the final state $\chi'$ has the same charge state as the initial state $\chi$, but is energetically excited compared to the initial state. To obtain the analytic expression of the diagram, we can follow the same prescription as in section E, or simply use the diagrammatic rules derived in Ref. \cite{gov08}. Thus we get
 \begin{eqnarray}
  \left(\hat K_\textsuperscript{IC}\right)^{\chi\chi}_{\chi'\chi'}&\equiv& \sum_{\vec k\sigma,\vec k'\sigma'}\sum_{mm'\nu}\int^\infty_0 dt'\,e^{-0^+t'}\int^{t'}_0 dt'_1 \int^{t'_1}_0 dt'_2 \nonumber \\*
&& \times\left\langle\hat c^\dagger_{S\vec k\sigma}(\tau)\hat c_{S\vec k\sigma}(t_2)\right \rangle\left\langle\hat c_{D\vec k'\sigma'}(t_1)\hat c^\dagger_{D\vec k'\sigma'}(t)\right \rangle\frac{|T_S|^2|T_D|^2}{\hbar^4} \nonumber \\*
&&\times\langle\chi'|\hat d_{m'\sigma'}(t)|\nu\rangle\langle\nu|\hat d^\dagger_{m\sigma}(t_2)|\chi\rangle\langle\chi|\hat d_{m\sigma}(\tau)|\nu\rangle \langle\nu|\hat d^\dagger_{m'\sigma'}(t_1)|\chi'\rangle \nonumber \\*
&=&-i\hbar\Gamma_S\Gamma_D \sum_\nu\int \frac{d\omega}{2\pi}\frac{d\omega'}{2\pi}D_S(\omega,\Delta)D_D(\omega',\Delta) \nonumber \\*
&&\times \frac{f_S(\omega)(1-f_D(\omega'))}{(-\omega+E_{\nu}-E_\chi+i0^+)(\omega'-\omega+E_{\chi'}-E_{\chi}+i0^+)} \nonumber \\*
&&\times\frac{1}{\omega'+E_{\chi'}-E_\nu+i0^+}. \nonumber
 \end{eqnarray}
The analysis of the kernel component and its remarkable contributions is done in the same way as depicted before for the elastic cotunneling case. Again we first focus on the low temperature regime and derive the condition for the standard inelastic cotunneling events in the Coulomb blockade region. We only consider the case in which the diagonal zeros of the denominator touch the corner of the blue region in Fig.~\ref{fig:condecot1}(b). In that case the condition for the bias threshold results in $\Delta+\mu_D=\omega'=\omega+E_\chi-E_{\chi'}=-\Delta+\mu_S+E_\chi-E_{\chi'}$. If we further use the energy difference $E_{\chi'}-E_\chi=\delta$ between the states of the CNT, we obtain the onset of the inelastic cotunneling peak in the current-voltage characteristics at bias voltage $|V_\textsuperscript{SD}|=(2\Delta+\delta)/e$. For higher temperatures an additional situation has to be considered. For thermally excited transport features we investigate the case when the diagonal zeros hit the orange regions in Fig.
~\ref{fig:condecot2}(b). Here we can give an additional requirement, $-\Delta+\mu_D=\omega'=\omega+E_\chi-E_{\chi'}=-\Delta+\mu_S+E_\chi-E_{\chi'}$, resulting in an onset of a thermal dependent peak in the conductance measurements at bias voltage $|V_\textsuperscript{SD}|=\delta/e$. The peak height of the thermal replica of the standard inelastic cotunneling grows with increasing temperature as more quasiparticles occupy the excited states and thus can contribute to the corresponding transport processes.

A quantitative approximation for the inelastic cotunneling rates in the middle of the Coulomb diamond for the process shown in Fig.~\ref{fig:recot}(b) can now be obtained when we investigate the integrand in the energy region $\omega'\approx\omega-\delta$. In the same manner as before we can then write the rate in Coulomb diamond \textcircled{\bf\footnotesize{3}} as
\begin{eqnarray}
 \Gamma_\textsuperscript{EC}^{3\to3^*}&\approx&-\hbar\int \frac{d\omega}{2\pi}D_S(\omega,\Delta)D_D(\omega-\delta,\Delta) \nonumber \\*
&& \times\frac{\Gamma_S\Gamma_D f_S(\omega)(1-f_D(\omega-\delta))}{(-\omega+E_4-E^0_3)(\omega-\delta+E^*_3-E_4)} \nonumber \\*
&\approx& \hbar\left(\frac{2}{U}\right)^2\Gamma_S\Gamma_D\int \frac{d\omega}{2\pi}D_S(\omega,\Delta)D_D(\omega-\delta,\Delta)f_S(\omega)[1-f_D(\omega-\delta)],\nonumber
\end{eqnarray}
where the 3-particle energy $E^*_3$ of the excited state was used.

\section*{References}

\bibliography{letter}

\end{document}

%% file: authors.tex
\author{S~Ratz$^1$, A~Donarini$^1$, D~Steininger$^2$, T~Geiger$^2$, A~Kumar$^2$\footnote{Deceased 23rd July 2014}, A~K~H\"uttel$^2$, Ch~Strunk$^2$ and M~Grifoni$^1$} 
\address{$^1$ Institute for Theoretical Physics, University of Regensburg, 93040 Regensburg, Germany}
\address{$^2$ Institute for Experimental and Applied Physics, University of Regensburg, 93040 Regensburg, Germany}
\ead{sascha.ratz@ur.de}

%% file: acknowledgement.tex
The authors acknowledge fruitful discussions with C. Chapelier. We thankfully acknowledge the support from the Deutsche Forschungsgemeinschaft (DFG) within GRK 1570, SFB 689, Emmy Noether (Hu 1808/1) and the EU FP7 Project SE2ND.

%% file: main1_1.bbl
\providecommand{\newblock}{}
\begin{thebibliography}{10}
\expandafter\ifx\csname url\endcsname\relax
  \def\url#1{{\tt #1}}\fi
\expandafter\ifx\csname urlprefix\endcsname\relax\def\urlprefix{URL }\fi
\providecommand{\eprint}[2][]{\url{#2}}

\bibitem{glaz89}
Glazman L~I and Matveev K~A 1989 {\em JETP Lett.\/} {\bf 49} 659

\bibitem{bas99}
Baselmans J, Morpurgo A~F, van Wees B and Klapwijk T~M 1999 {\em Nature\/} {\bf
  397} 43

\bibitem{roz01}
Rozhkov A~V, Arovas D~P and Guinea F 2001 {\em Phys. Rev. B\/} {\bf 64}(23)
  233301

\bibitem{Doh05}
Doh Y~J, van Dam J~A, Roest A~L, Bakkers E~P~A~M, Kouwenhoven L~P and {De
  Franceschi} S 2005 {\em Science\/} {\bf 309} 272

\bibitem{dam06}
van Dam J~A, Nazarov Y~V, Bakkers E, {De Franceschi} S and Kouwenhoven L 2006
  {\em Nature\/} {\bf 442} 667

\bibitem{Jar06}
Jarillo-Herrero P, van Dam J~A and Kouwenhoven L 2006 {\em Nature\/} {\bf 439}
  953

\bibitem{Sch01}
Scheer E, Belzig W, Naveh Y, Devoret M~H, Esteve D and Urbina C 2001 {\em Phys.
  Rev. Lett.\/} {\bf 86}(2) 284

\bibitem{Bui03}
Buitelaar M~R, Belzig W, Nussbaumer T, {Babi\ifmmode \acute{c}\else {\'c}\fi{}}
  B, Bruder C and Sch{\"o}nenberger C 2003 {\em Phys. Rev. Lett.\/} {\bf 91}(5)
  057005

\bibitem{andersen11}
Andersen B~M, Flensberg K, Koerting V and Paaske J 2011 {\em Phys. Rev.
  Lett.\/} {\bf 107} 256802

\bibitem{deon11}
Deon F, Pellegrini V, Giazotto F, Biasiol G, Sorba L and Beltram F 2011 {\em
  Phys. Rev. B\/} {\bf 84}(10) 100506

\bibitem{levy97}
Yeyati A~L, Cuevas J~C, L{\'o}pez-D{\'a}valos A and Mart{\'i}n-Rodero A 1997
  {\em Phys. Rev. B\/} {\bf 55}(10) 6137

\bibitem{GoLo04}
Golovach V~N and Loss D 2004 {\em Phys. Rev. B\/} {\bf 69}(24) 245327

\bibitem{eich07}
Eichler A, Weiss M, Oberholzer S, Sch{\"o}nenberger C, {Levy Yeyati} A, Cuevas
  J~C and Mart{\'i}n-Rodero A 2007 {\em Phys. Rev. Lett.\/} {\bf 99}(12) 126602

\bibitem{grove09}
Grove-Rasmussen K, J{\o}rgensen H~I, Andersen B~M, Paaske J, Jespersen T~S,
  Nyg{\aa}rd J, Flensberg K and Lindelof P~E 2009 {\em Phys. Rev. B\/} {\bf
  79}(13) 134518

\bibitem{dirk09}
Dirks T, Chen Y~F, Birge N~O and Mason N 2009 {\em Applied Physics Letters\/}
  {\bf 95} 192103

\bibitem{fran10}
{De Franceschi} S, Kouwenhoven L, Schonenberger C and Wernsdorfer W 2010 {\em
  Nat. Nano\/} {\bf 5} 703

\bibitem{Pfaller13}
Pfaller S, Donarini A and Grifoni M 2013 {\em Phys. Rev. B\/} {\bf 87}(15)
  155439

\bibitem{Gaass14}
Gaass M, Pfaller S, Geiger T, Donarini A, Grifoni M, H{\"u}ttel A~K and Strunk
  C 2014 {\em Phys. Rev. B\/} {\bf 89}(24) 241405

\bibitem{joh99}
Johansson G, Bratus E~N, Shumeiko V~S and Wendin G 1999 {\em Phys. Rev. B\/}
  {\bf 60}(2) 1382

\bibitem{gunel12}
G{\"u}nel H~Y, Batov I~E, Hardtdegen H, Sladek K, Winden A, Weis K, Panaitov G,
  Gr{\"u}tzmacher D and Sch{\"a}pers T 2012 {\em Journal of Applied Physics\/}
  {\bf 112} 034316

\bibitem{averin90}
Averin D~V and Nazarov Y~V 1990 {\em Phys. Rev. Lett.\/} {\bf 65}(19) 2446

\bibitem{kong98}
Kong J, Soh H, Cassell A, Quate C and Dai H 1998 {\em Nature\/} {\bf 395}(6705)
  878

\bibitem{pall08}
Pallecchi E, Gaa{\ss} M, Ryndyk D~A and Strunk C 2008 {\em Applied Physics
  Letters\/} {\bf 93} 072501

\bibitem{mart89}
Martinis J~M and Kautz R~L 1989 {\em Phys. Rev. Lett.\/} {\bf 63}(14) 1507

\bibitem{Note1}
The discrepancy in gate voltage range between Fig.~\ref {fig:sampleA}(b) and
  Figs.~\ref {fig:sampleA}(c),(d) is caused by a long-time scale drift of all
  Coulomb blockade features. Sequential tunneling features of this data set
  have already been discussed in Ref. \cite {Gaass14}

\bibitem{kulm72}
Hulm J, Jones C, Hein R and Gibson J 1972 {\em J. Low Temp. Phys.\/} {\bf 7}
  291

\bibitem{may72}
Mayadas A~F, Laibowitz R~B and Cuomo J~J 1972 {\em J. Appl. Phys.\/} {\bf 43}
  1287

\bibitem{Kum14}
Kumar A, Gaim M, Steininger D, Yeyati A~L, Mart{\'i}n-Rodero A, H{\"u}ttel A~K
  and Strunk C 2014 {\em Phys. Rev. B\/} {\bf 89}(7) 075428

\bibitem{bui02}
Buitelaar M~R, Nussbaumer T and Sch{\"o}nenberger C 2002 {\em Phys. Rev.
  Lett.\/} {\bf 89}(25) 256801

\bibitem{sia04}
Siano F and Egger R 2004 {\em Phys. Rev. Lett.\/} {\bf 93}(4) 047002

\bibitem{Cle06}
Cleuziou J~P, Wernsdorfer W, Bouchiat V, Ondarcuhu T and Monthioux M 2006 {\em
  Nat. Nano\/} {\bf 1} 53

\bibitem{Kim13}
Kim B~K, Ahn Y~H, Kim J~J, Choi M~S, Bae M~H, Kang K, Lim J~S, L{\'o}pez R and
  Kim N 2013 {\em Phys. Rev. Lett.\/} {\bf 110}(7) 076803

\bibitem{Lee14}
{Lee Eduardo} J~H, Jiang X, Houzet M, Aguado R, Lieber C~M and {De Franceschi}
  S 2014 {\em Nat. Nano\/} {\bf 9} 79

\bibitem{Chang13}
Chang W, Manucharyan V~E, Jespersen T~S, Nyg{\aa}rd J and Marcus C~M 2013 {\em
  Phys. Rev. Lett.\/} {\bf 110}(21) 217005

\bibitem{laird14}
{Laird} E~A, {Kuemmeth} F, {Steele} G, {Grove-Rasmussen} K, {Nyg{\aa}rd} J,
  {Flensberg} K and {Kouwenhoven} L~P 2014 {\em arxiv:1403.6113\/} (12)

\bibitem{Bog58}
Bogoljubov N 1958 {\em Il Nuovo Cimento\/} {\bf 7} 794

\bibitem{Val58}
Valatin J 1958 {\em Il Nuovo Cimento\/} {\bf 7} 843

\bibitem{Jos62}
Josephson B 1962 {\em Physics Letters\/} {\bf 1} 251

\bibitem{Bar62}
Bardeen J 1962 {\em Phys. Rev. Lett.\/} {\bf 9}(4) 147--149

\bibitem{Blu12}
Blum K 2012 {\em {Density Matrix Theory and Applications}\/} {Springer Series
  on Atomic} (Springer)

\bibitem{weymann05}
Weymann I, K{\"o}nig J, Martinek J, {Barna\ifmmode \acute{s}\else {\'s}\fi{}} J
  and Sch{\"o}n G 2005 {\em Phys. Rev. B\/} {\bf 72}(11) 115334

\bibitem{Koller10}
Koller S, Grifoni M, Leijnse M and Wegewijs M~R 2010 {\em Phys. Rev. B\/} {\bf
  82}(23) 235307

\bibitem{gov08}
Governale M, Pala M~G and K{\"o}nig J 2008 {\em Phys. Rev. B\/} {\bf 77}(13)
  134513

\bibitem{Note2}
We replace the Heaviside function $\Theta (|\omega |-\Delta )\to \frac {1}{\exp
  (\gamma ^{-1}(\omega +\Delta ))+1}+\frac {1}{\exp (\gamma ^{-1}(-\omega
  +\Delta ))+1}$ by a blurred step function. Despite $\gamma $ is introduced
  empirically in this work, it can be shown that higher order processes
  involving quasiparticles lead to level broadening in the quantum dot and thus
  also to regularization of the divergence caused by the BCS density of
  states~\cite {levy97} similar to that provided by $\gamma $ here.

\bibitem{Dynes78}
Dynes R~C, Narayanamurti V and Garno J~P 1978 {\em Phys. Rev. Lett.\/} {\bf
  41}(21) 1509

\bibitem{jesp11}
{Jespersen T S}, {Grove-Rasmussen K}, {Paaske J}, {Muraki K}, {Fujisawa T},
  {Nygard J} and {Flensberg K} 2011 {\em Nat. Phys.\/} {\bf 7} 348

\bibitem{peck13}
{Pecker S}, {Kuemmeth F}, {Secchi A}, {Rontani M}, {Ralph D C}, {McEuen P L}
  and {Ilani S} 2013 {\em Nat. Phys.\/} {\bf 9} 576

\bibitem{schl05}
Schleser R, Ihn T, Ruh E, Ensslin K, Tews M, Pfannkuche D, Driscoll D~C and
  Gossard A~C 2005 {\em Phys. Rev. Lett.\/} {\bf 94}(20) 206805

\bibitem{huttel09}
H{\"u}ttel A~K, Witkamp B, Leijnse M, Wegewijs M~R and van~der Zant H~S~J 2009
  {\em Phys. Rev. Lett.\/} {\bf 102}(22) 225501

\bibitem{ander63}
Anderson P~W and Rowell J~M 1963 {\em Phys. Rev. Lett.\/} {\bf 10}(6) 230

\bibitem{jos74}
Josephson B~D 1974 {\em Rev. Mod. Phys.\/} {\bf 46}(2) 251

\bibitem{BarCooSch57}
Bardeen J, Cooper L~N and Schrieffer J~R 1957 {\em Phys. Rev.\/} {\bf 108}(5)
  1175

\bibitem{Sch99}
Schrieffer J 1999 {\em {Theory of Superconductivity}\/} {Advanced Book Program
  Series} (Advanced Book Program, Perseus Books)

\end{thebibliography}
